%% file: Main_TMC.tex
\documentclass[lettersize, jounal]{IEEEtran}
\usepackage{amsmath,amsfonts}
\usepackage{algorithmic}
\usepackage{array}
\usepackage[caption=false,font=normalsize,labelfont=sf,textfont=sf]{subfig}
\usepackage{textcomp}
\usepackage{stfloats}
\usepackage{url}
\usepackage{verbatim}
\usepackage{xcolor}
\usepackage{rotating}
\usepackage[T1]{fontenc}

\usepackage{graphicx}
\def\BibTeX{{\rm B\kern-.05em{\sc i\kern-.025em b}\kern-.08em
    T\kern-.1667em\lower.7ex\hbox{E}\kern-.125emX}}
\usepackage{balance}

\begin{document}

\title{Detecting and Characterising Mobile App Metamorphosis in Google Play Store}

\author{D. Denipitiyage, B. Silva, K. Gunathilaka, S. Seneviratne, A. Mahanti, A. Seneviratne, S. Chawla
\thanks{D. Denipitiyage, B. Silva, S. Seneviratne and A. Mahanti are with the School of Computer science, University of Sydney, Australia (e-mail: dden5444@uni.sydney.edu.au; bpin9254@uni.sydney.edu.au; suranga.seneviratne@sydney.edu.au; anirban.mahanti@sydney.edu.au)}
\thanks{K. Gunathilaka is with the Department of Computer Science and Engineering, University of Moratuwa, Sri Lanka (email: gihanthakavishka@gmail.com)}
\thanks{A. Seneviratne is with the University of New South Wales (UNSW), Sydney, Australia (e-mail: a.seneviratne@unsw.edu.au) }
\thanks{S. Chawla is with the Qatar Computing Research Institute, Hamad Bin Khalifa University (HBKU) (e-mail: schawla@hbku.edu.qa)}}

\markboth{IEEE TRANSACTIONS ON MOBILE COMPUTING,~Vol.~24, No.8, August~2025}{How to Use the IEEEtran \LaTeX\ Templates}%
\maketitle

\begin{abstract}
App markets have evolved into highly competitive and dynamic environments for developers. While the traditional app life cycle involves incremental updates for feature enhancements and issue resolution, some apps deviate from this norm by undergoing significant transformations in their use cases or market positioning. We define this previously unstudied phenomenon as `app metamorphosis.' 

In this paper, we propose a novel and efficient multi-modal search methodology to identify apps undergoing metamorphosis and apply it to analyse two snapshots of the Google Play Store taken five years apart. Our methodology uncovers various metamorphosis scenarios, including re-births, re-branding, re-purposing, and others, enabling comprehensive characterisation. Although these transformations may register as successful for app developers based on our defined success score metric (e.g., re-branded apps performing approximately 11.3\% better than an average top app), we shed light on the concealed security and privacy risks that lurk within, potentially impacting even tech-savvy end-users.
\end{abstract}




\input sections/1_Introduction
\input sections/2_Related

\input sections/3_Datasets
\input sections/4_Methodology
\input sections/5_Results

\input sections/6_ResultsAnalysis

\input sections/7_Discussion
\input sections/8_Conclusion

\section{Acknowledgment}
This research was supported by the Australian Government through the Australian Research Council’s Discovery Projects funding scheme (Project ID DP220102520).

\bibliographystyle{ieeetr}
\bibliography{biblio}

\vspace{-30mm}

\begin{IEEEbiography}[{\includegraphics[width=1in,height=1.25in,clip,keepaspectratio]{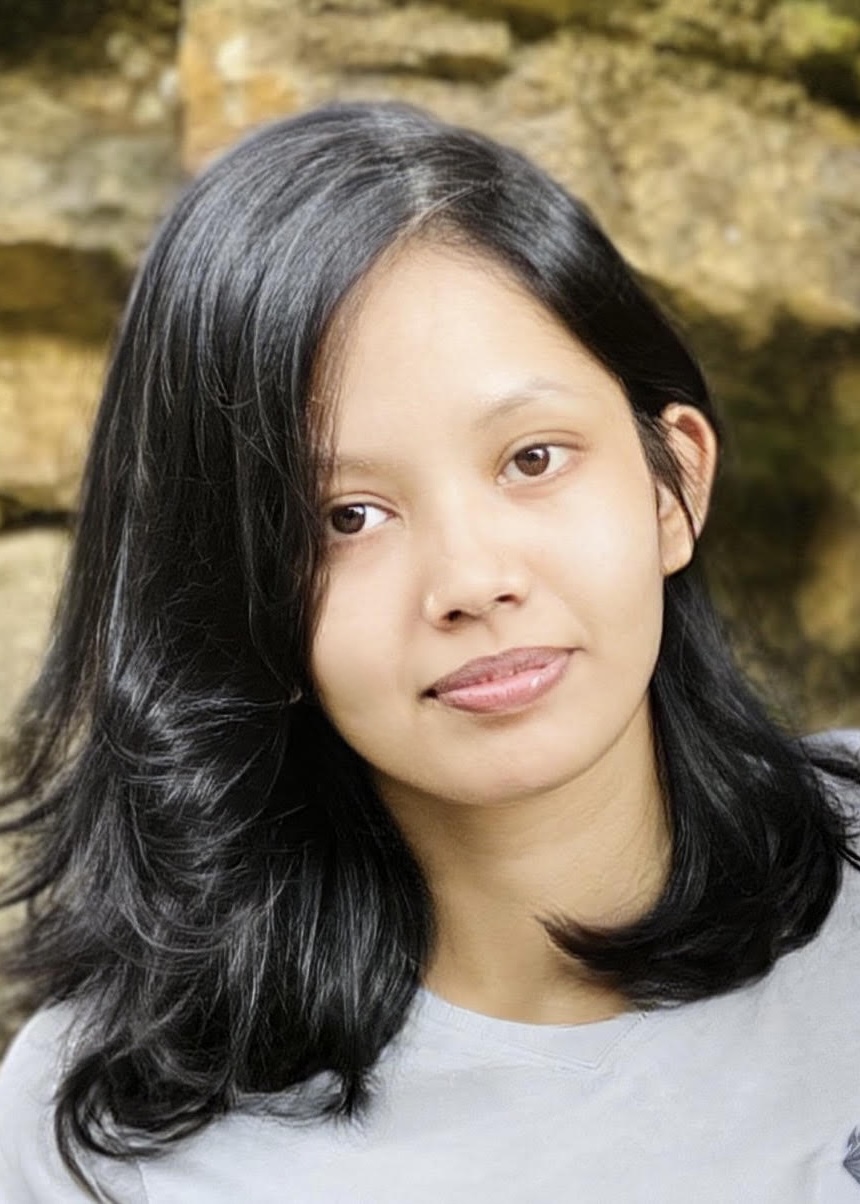}}]{Dishanika Denipitiyage}
	received her Bachelors degree in Electronic and Telecommunication Engineering from University of Moratuwa, Sri Lanka in 2020. She is currently working toward the PhD degree with the School of Computer Science, University of Sydney, Australia. She worked as a Senior software engineear at SenzMate (Pvt) Ltd, Sri Lanka in 2022 and as a Visiting Research Intern at Singapore University of Technology and Design (SUTD), Singapore in 2017. Her research interests include Self-Supervised Learning and Multi-Modal learning.
\end{IEEEbiography}

\vskip -3\baselineskip plus -1fil
\vspace{8mm}

\begin{IEEEbiography}[{\includegraphics[width=1in,height=1.25in,clip,keepaspectratio]{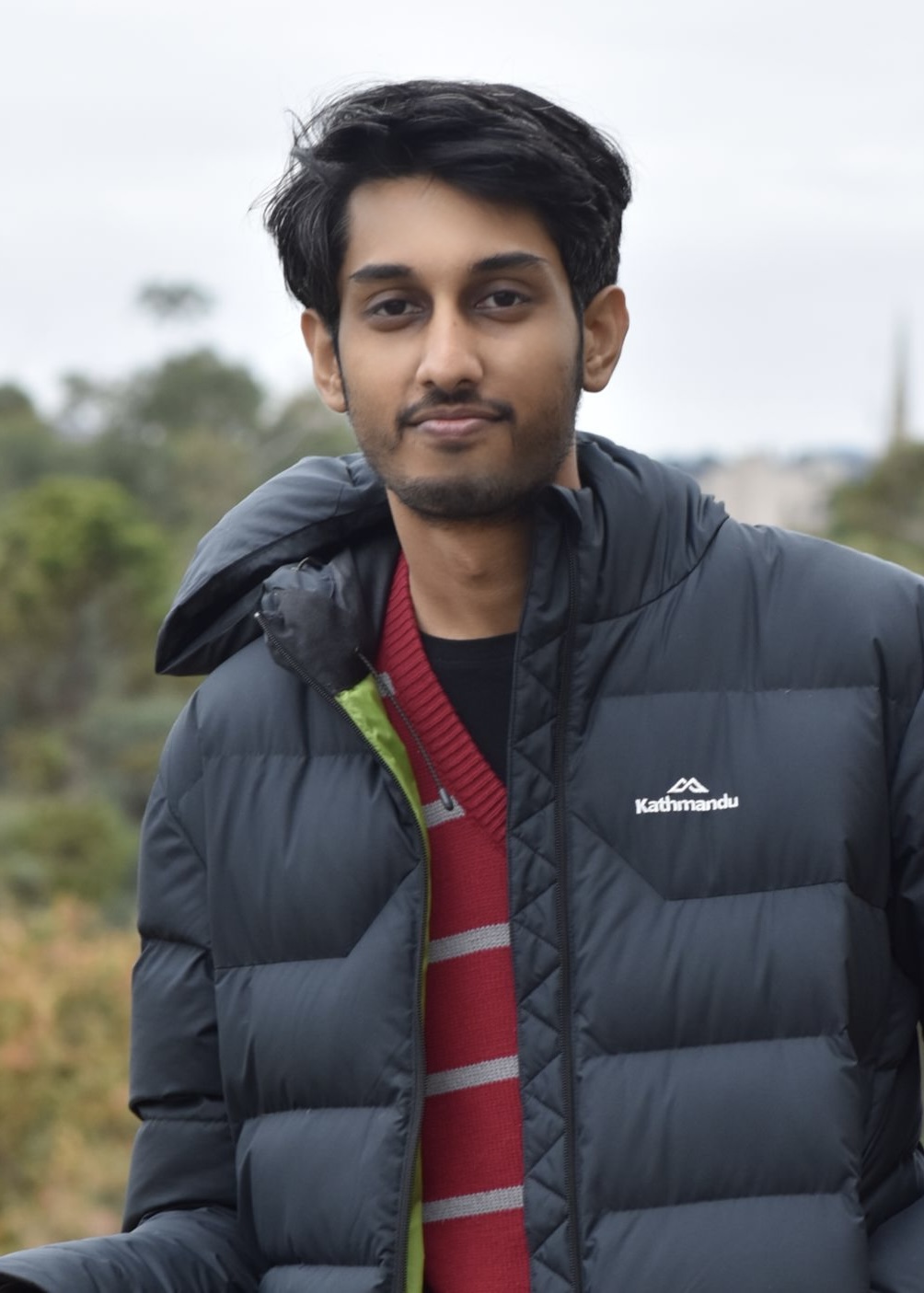}}]{Bhanuka Silva}
received his Bachelors degree in Electronic and Telecommunication Engineering (First Class Hons.) from University of Moratuwa, Sri Lanka in 2020. He also worked as a Visiting Research Intern at Data61-CSIRO, Brisbane in 2018 and is currently a doctoral student at the University of Sydney and his current research focuses on conducting privacy compliance checks in mobile app eco-systems by leveraging state-of-the art natural language processing techniques.
\end{IEEEbiography}

\vskip -3\baselineskip plus -1fil
\vspace{8mm}

\begin{IEEEbiography}[{\includegraphics[width=1in,height=1.25in,clip,keepaspectratio]{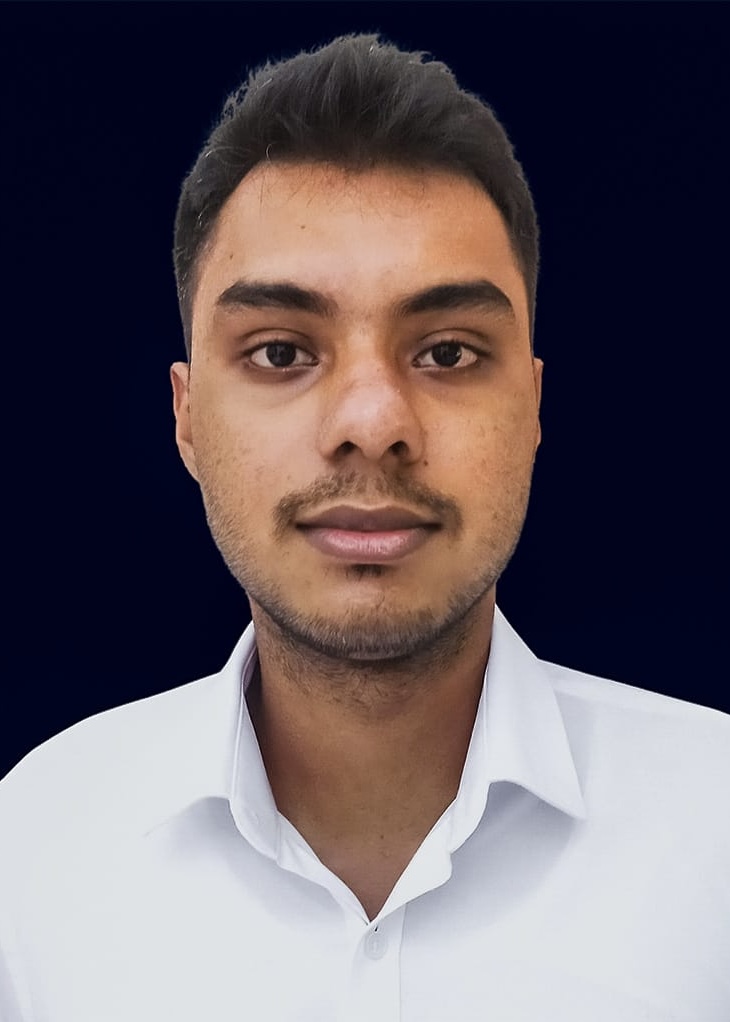}}]{Kavishka Gunathilaka}
is a B.Sc. Eng. undergraduate in the Department of Computer Science and Engineering of the University of Moratuwa, Sri Lanka. He also worked as a Research Affiliate at the University of Sydney in 2023. His primary research interests include machine learning, data 
science, and artificial intelligence.
\end{IEEEbiography}

\vskip -3\baselineskip plus -1fil
\vspace{8mm}

\begin{IEEEbiography}[{\includegraphics[width=1in,height=1.25in,clip,keepaspectratio]{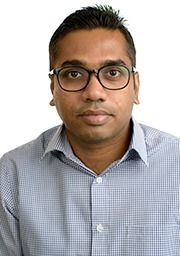}}]{Suranga Seneviratne}
is a Senior Lecturer in Security at the School of Computer Science, The University of Sydney. He received his Ph.D. from the University of New South Wales, Australia in 2015. His current research interests include privacy and security in mobile systems, AI applications in security, and behavior biometrics. Before moving into research, he worked nearly six years in the telecommunications industry in core network planning and operations. He received his bachelor degree from University of Moratuwa, Sri Lanka in 2005.
\end{IEEEbiography}

\vskip -2\baselineskip plus -1fil
\vspace{8mm}

\begin{IEEEbiography}[{\includegraphics[width=1in,height=1.25in,clip,keepaspectratio]{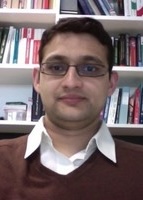}}]{Anirban Mahanti}'s technical expertise is at the intersection of computer networking, data science, and machine learning. Anirban earned his Ph.D. in Computer Science from the University of Saskatchewan, Canada, in 2003, his MSc in Computer Science in 1999, and his BE in Computer Science and Engineering from the Birla Institute of Technology, India, in 1993. He is currently an Honorary Senior Research Fellow at the University of Sydney.
\end{IEEEbiography}
\vskip -2\baselineskip plus -1fil
\begin{IEEEbiography}[{\includegraphics[width=1in,height=1.25in,clip,keepaspectratio]{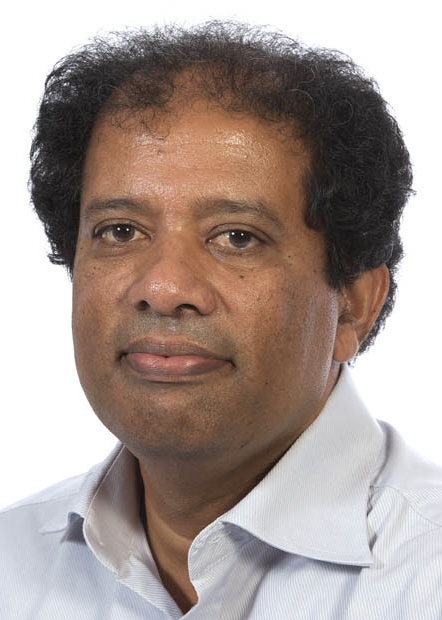}}]{Aruna Seneviratne}
(Senior Member, IEEE) is currently a foundation professor of telecommunications with the University of New South Wales, Sydney, Australia, where he holds the Mahanakorn chair of telecommunications. He was with a number of other universities in Australia, UK, and France, as well as industrial organizations, including Muirhead, Standard Telecommunication Labs, Avaya Labs, and Telecom Australia (Telstra). He held visiting appointments with INRIA, France. His research inter- ests include physical analytics, technologies that enable applications to interact intelligently and securely with their environment in real time. Recently, his team has been working on using these technologies in behavioural biometrics, optimizing the performance of wearables, and IoT system verification. He was the recipient of several fellowships, including one at British Telecom and one at Telecom Australia Research Labs.
\end{IEEEbiography}

\vskip -2\baselineskip plus -1fil
\vspace{8mm}

\begin{IEEEbiography}[{\includegraphics[width=1in,height=1.25in,clip,keepaspectratio]{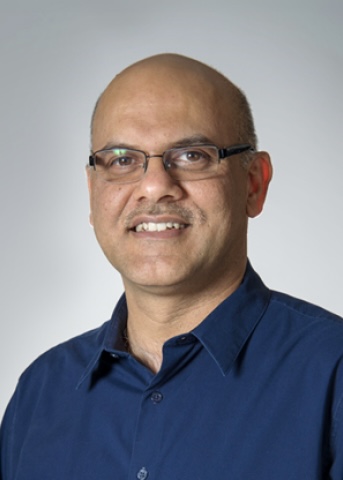}}]{Sanjay Chawla}
is Research Director of QCRI’s Data Analytics department. Prior to joining QCRI, Dr. Chawla was a Professor in the Faculty of Engineering and IT at the University of Sydney. From 2008-2011, he also served as the Head (Department Chair) of the School of Information Technologies. He was an academic visitor at Yahoo! Research in 2012. He received his PhD from the University of Tennessee (USA) in 1995. His research is in data mining and machine learning with a specialization in spatio-temporal data mining, outlier detection, class imbalanced classification, and adversarial learning. 
\end{IEEEbiography}

\newpage
\appendix
\input sections/9_Appendix

\end{document}

%% file: sections/1_Introduction.tex
\section{Introduction}
\label{sec:introduction}


The mobile app industry is highly competitive, with over 2.5 million apps on the Google Play Store~\cite{googleplay} and 1.5 million apps on the Apple App Store~\cite{appleapp}, it can be difficult for apps to stand out from the crowd. To be successful, app developers need to create innovative and user-friendly apps, update existing apps, and respond to new operating system features and market demands.

While the typical life cycle of an app involves releasing it to the app market and periodically updating it to add new features or fix bugs and security issues, some apps deviate from this pattern of incremental updates. Instead, they undergo dramatic changes in their use cases or market positioning. The methods to match apps between two different time spans and observe such dramatic transformations remain largely unexplored. We define this relatively unknown phenomenon as \emph{``app metamorphosis''} and there are multiple reasons and implications of it based on our findings.





For instance, sometimes developers may use the \textit{re-branding} strategy to uplift a struggling app. This re-branding will involve changing the app's name, logo, or overall look and feel. The aim is to improve the app's image or to make it more relevant to a new target audience. Re-branding can also happen after a merger or an acquisition, as in the case of the popular short video-sharing social networking app \textit{TikTok}, which used to be called  \textit{Musical.ly} with a slightly different purpose of sharing short lip-synced videos~\cite{TikTok}. 

Similarly, some developers looking to extend the useful life of their successful apps employ the \textit{re-purposing} strategy. Re-purposing an app means changing its core functionality or purpose. This enables the developers to change their app's core functionality while retaining its app audience, download numbers, ratings, etc. Equally, some developers may re-purpose their apps when their core functionalities become redundant due to them being added to the operating system as standard features. For example, due to hardware advances in smartphone cameras, some camera apps such as \emph{Noah Camera} became photo editors.



\begin{figure*}[t]
    \centering
    \includegraphics[width=0.98\textwidth]{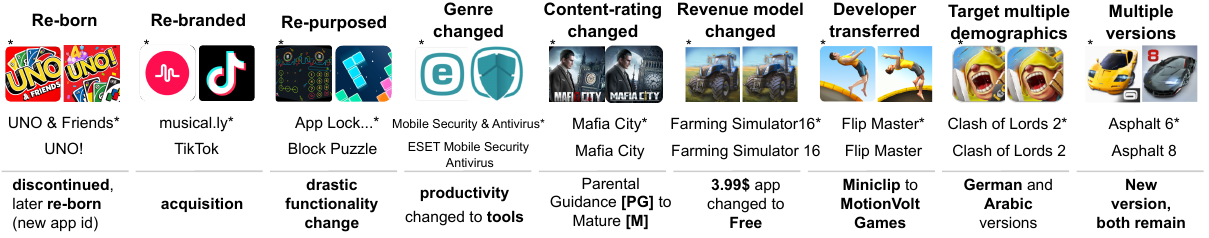}
    \caption{Identified app metamorphosis categories and a notable example pair for each category. (*: 2018 app)}
    \label{fig:re-emergence tree diagram}
\end{figure*}



Developers can also attempt to revive apps by \textit{re-birthing} them. That means re-introducing a previously discontinued or an about-to-change app under a new identity (i.e., a new app ID), with minimum or no functionality change compared to the original one. This is a major undertaking, but it can be a successful way to create a new and improved app that is more likely to succeed. As an example, once \emph{UNO \& Friends} was discontinued by Ubisoft, it was revived back by a different developer \emph{Mattel163} as \emph{UNO!} and became successful by reaching a wider audience. Another potentially adverse reason for \textit{re-birthing} is to overcome bans imposed by Google on developers. In Fig.~\ref{fig:re-emergence tree diagram} we provide a summary of these metamorphosis scenarios we identify.


To date, motivations for \textit{app metamorphosis} are not fully understood as there have been no systematic studies that focus on this aspect of the mobile app ecosystem.  An investigation of \textit{app metamorphosis} that focuses on the characteristics and prevalence of \textit{app metamorphosis} will provide further insights into how the mobile app ecosystem is evolving. Moreover, such a study will help app market operators to identify different forms of \textit{app metamorphosis} and their impact on the apps and enable developers to understand the implications of their actions. It is also important to understand how successful or unsuccessful the apps that underwent metamorphosis are. This will provide insights to the app developers to make decisions on app updates and monetisation strategies. Finally, there can be privacy and security implications because of these app transformations.


To this end, in this paper, we compare two snapshots of the Google Play Store that are five years apart and propose a search methodology to identify similar apps across datasets that will help identify apps that possibly went through a ``metamorphosis'' phase. Next, we provide a taxonomy and a set of rules to categorise various cases of ``metamorphosis'' identify and discuss reasons behind them. More specifically, we make the following contributions.

\begin{itemize}
\item We propose a multi-modal search methodology for identifying similar apps across datasets, incorporating dimensions such as app icon similarity, app description similarity, app name, and developer name similarity. Usually, app re-identification is straightforward, and you can use the \emph{app ID} to do that. However, in this case, we want something beyond that because part of our objective is to identify similar apps with different \emph{app IDs} between datasets.
\item We apply our search methodology to analyse two snapshots of the Google Play Store five years apart, focusing primarily on the top 10,000 apps in 2018, and identify apps that have possibly gone through ``metamorphosis'' between  2018-2023 and provide a taxonomy of various cases of ``metamorphosis'' and build a rule set to identify apps belonging to those cases.
\item We introduce a ``success score'' to quantitatively characterise how successful an app progressing from 2018 to 2023 is, compared to the natural growth in Android eco-system. Based on the overall results, we show that app \emph{re-branding} has the highest possibility of becoming successful ($\sim11.3\%$ better) followed up by \emph{re-purposing} ($\sim4.3\%$ better) while \emph{re-birthing} is the most challenging. 
\item We further discuss how metamorphosis can ascertain privacy and security risks, for example, an adverse developer could introduce a counterfeit app, disguising as a legitimate re-birth for discontinuing apps.
\end{itemize}

The rest of the paper is organised as follows. In Sec.~\ref{Sec:Related}, we present the related work, and in Sec.~\ref{Sec:Datasets} we present our datasets. Sec.~\ref{Sec:Methodology} presents our multi-modal app similarity search algorithm and Sec.~\ref{Sec:Results} presents its performance in controlled experiments. Sec.~\ref{Sec:Analysis} presents the results and findings of using our search methods to identify apps demonstrating ``metamorphosis'' between 2018 to 2023. Sec.~\ref{sec:sec_security} discusses the privacy and security implications of our findings. Finally, Sec.~\ref{Sec:Conclusion} concludes the paper.



%% file: sections/2_Related.tex
\section{Related Work}
\label{Sec:Related}

Various empirical studies have been conducted on Google Play Store and other app stores focusing on different aspects such as; general statistics related to apps~\cite{viennot2014measurement,wang2018beyond}, advertising and analytics library behaviours~\cite{chen2016following, riganelli2022proactive}, characterising and detecting malware and malpractices~\cite{wu2021android, wang2018automated, malware_grace2012riskranker, malware_shabtai2012andromaly, malware_wu2012droidmat, malware_yuan2014droid}, and privacy and security analysis of one specific type of apps~\cite{ikram2016analysis, nguyen2021share}. However, only a limited amount of work focused on studying the  evolution of apps over a longer time period. \\ \vspace{-3mm}


\noindent{\textbf{App life cycle in Google Play -}} The usual life cycle of an app in Google Play Store involves the developer publishing the app by selecting some of the metadata such as category, content rating, and self-reporting data collection and sharing practises~\cite{hageman2023mixed, nguyen2021share}. Once it is published, the app will be updated periodically to release new features or fix bugs. The vast majority of apps will continue to be like this; however, some apps may get discontinued (i.e., no further updates), removed from Google Play Store (i.e., either developer decides to remove the app, or Google removes the app or ban the developer for various reasons~\cite{seneviratne2015early}) or go through ``metamorphosis'' - which is the focus of this paper. 

To date, the app life cycle has been characterised predominantly from an app update point of view. For instance, Pothuraju et al.~\cite{updates_potharaju2017longitudinal} claim that nearly 76\% of apps did not get any update in Play Store within their monitoring dataset for a period of approximately six months while a minority got nearly hundreds of updates which may point to newly released apps that could require many bug fixes.  Though these updates are supposed to fix bugs or provide better security,  Moller et al.~\cite{updates_moller2012update} noted that even after a week of an update, app users still tend to use the older version. Vincent et at.~\cite{taylor2017update} used PlayDrone~\cite{viennot2014measurement} to extract permission usage of apps and observed that popular apps request additional dangerous permissions with subsequent updates. 



Another body of work checked how some individual app libraries, such as advertising and analytic libraries, are being updated~\cite{ahasanuzzaman2020longitudinal, wang2019understanding}. However, in all these cases, the apps do not go through ``metamorphosis'', and app developers remain the same. We also note that the app life cycle we consider in this paper is different from the app life cycle considered by some works ~\cite{long_term_evolution_2023,vagrani2017decline}, which focus on the app usage point of view, such as when a  user installs an app, and abandon it after using it for a period of time. \\ \vspace{-3mm} 



\noindent{\textbf{Disguising (or misleading) app users - }} Beyond malware, multiple works explored various disguises and malpractices happening in Google Play Store. One of the common ways to distribute malware in the mobile ecosystem is through repackaging of legitimate apps. This process includes app downloading, de-compiling, manipulating, recompiling an application, and publishing it again in an app store. These repackaged apps are often distributed through third-party or unofficial app markets but can also appear on official app stores like Google Play under new app IDs. Several works looked into detecting repackaged apps by analysing bytecode similarity~\cite{hanna2013juxtapp, zhou2012detecting, zhang2014viewdroid, khan2020repacked, guan2016semantics}, app metadata such as APIs, permissions, and description~\cite{kywe2015detecting, zhou2013fast, ali2015opseq}, or runtime behavior monitoring~\cite{zhou2013appink}. While these work revealed privacy issues attached to mobile apps, our approach contributes differently, to a novel perspective of app life cycle changes. However, such work is related to ours because when detecting ``app metamorphosis'', we can incur some of these cases as false positives. 
Karunanayake et al.~\cite{karunanayake2020multi} proposed a multi-modal neural framework to identify counterfeit apps that impersonate popular apps. In contrast, Viennot et al.~\cite{viennot2014measurement} used clusters of similar apps, developer names, and certificate information to identify potential legitimate re-branding. As we present later in the results, some of the ``app metamorphosis'' cases we identify will have some overlaps with counterfeiting. Other forms of malpractices that have been investigated include; miscategorisation~\cite{miscategory_surian2017app} and spamming~\cite{seneviratne2015early, seneviratne2017spam}.

 \textit{To the best of our knowledge, ``app metamorphosis'' hasn't been studied or characterised to date, and ours is the first study that investigates this phenomenon by comparing two large snapshots in Google Play Store that are five years apart.}

%% file: sections/3_Datasets.tex
\section{Datasets}
\label{Sec:Datasets}

We use two main datasets in our work. These are two snapshots of the Google Play Store, collected in 2018 and 2023, respectively. They contain app metadata (e.g., app ID, app name, app category/genre, app description, developer name, etc.), app creatives (e.g., app icons and app screenshots), and for free apps, APK app executables. The 2018 dataset that was collected as a part of our previous work~\cite{karunanayake2020multi} contains metadata of over one million apps that were collected between January and March 2018. The 2023 dataset contains metadata of 1,280,142 apps and was collected between January and November 2023.  The data was collected using a Python-based crawler that started from an initial seed of apps, sourced by various top-lists in the Google Play Store progressively discovering new ones. To avoid any disturbances to the app market operation, we conducted our crawl conservatively at a low rate. \\ \vspace{-3mm}

\noindent{{\bf i) Top-k apps in 2018 and 2023 datasets -} Our key objective in this work is to identify apps that went through notable re-transformations between the two snapshots. To get better insights and to avoid low-quality apps affecting our results as noise, we focus only on the top-k apps in the 2018 dataset and their relevant counterpart in 2023. 
To identify the top-k apps in each dataset, we use the same method proposed by~\cite{seneviratne2015early} in sorting the datasets. That is, we sort the set of crawled apps in a dataset (either the 2018 dataset or the 2023 dataset), first based on the number of downloads, second based on rating count, and third based on the average stars and select the first k apps. Next, we constructed a `validation-set' and a `test-set' to fine-tune hyper-parameters of the app similarity search algorithm and to evaluate the performance. We discuss our search algorithm later in the Sec.~\ref{Sec:Methodology}}. \\ \vspace{-3mm}




\begin{figure}[ht]
    \centering
    \includegraphics[width=0.45\textwidth]{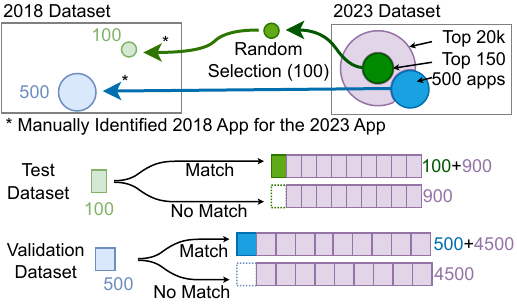}
    \caption{Creation of validation and test sets from 2018 and 2023 datasets.}
    \label{fig:dataset-structure}
\end{figure}

\begin{figure*}[t]
    \centering
    \includegraphics[width=0.99\textwidth]{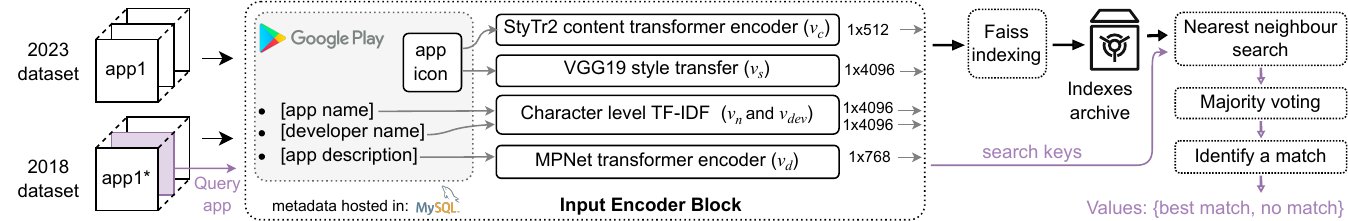}
    \caption{Overall methodology for query, key, and value based best match (if it exists - as emphasised in the purple colour path) retrieval  for counterpart app identification to an app ID query from 2018 dataset to be matched with indexed 2023 dataset. We utilise Faisis library for creating the indexes archive.}
    \label{fig:methodology}
\end{figure*}




\noindent{{\bf ii) Validation-sets}} - 
When we are searching for a counterpart app between the two datasets, there may be or may not be a matching app. For example, an app may have been discontinued.
To properly capture this idea, there must be ground-truth sets on \emph{matches} and \emph{no matches} to fine-tune our algorithm; therefore, we create two validation-sets. 

First, we create a validation-set for the \emph{match} scenario. As the first step, we select 500 apps from 1,280,142 apps, which has a counterpart in 2023 dataset (teal colour circle in Fig. 2). We exclude top-150 apps from the 2023 dataset when selecting these 500 apps for test-set creation. We search the counterpart 2018 app, first using the app ID and then manually verifying each one of them (blue colour arrow). The verification is done by cross-referencing the apps’ metadata between the two datasets. The figure in Appendix.~C shows how the top 500 apps are distributed across the 2023 dataset of 1.3M apps. As the next step, we add 4,500 more apps to the selected 500 apps from 2023 dataset, randomly selected from the top-20,000 apps of the 2023 dataset but excluding the first top-150 apps and 199 apps that are selected for the validation set (199 apps are in between top-150 and top-20,000). Now we have 500 apps from 2018 (light-blue colour circle) and 5000 apps from 2023 dataset with 500 guaranteed matching apps between
them. Our expectation is to identify all of them as matches during the validation step of our similarity search. To create the \emph{no-match} validation-set, we simply remove the 500 apps from 2023 side that were mapped to the 2018 dataset.\\ \vspace{-3mm}

\noindent{{\bf iii) Test-set} - 
To test the performance of our app similarity search algorithm with \emph{matching} and \emph{no-matching} apps, we create two controlled test-sets, quite similar to the method discussed before. We first randomly select 100 apps from top-150 apps in 2023 dataset (green colour circle in Fig.~\ref{fig:dataset-structure}). Then we select their counterpart 2018 app using the app ID followed up by a manual verification (light-green colour arrow). In total, we found that 117 app IDs from the top-150 apps in the 2023 dataset had exact matches in the 2018 dataset. Next, we add 900 more apps to the 2023 side by randomly selecting from the top-20,000 apps but excluding 199 apps of validation set in the 2023 dataset (purple colour ring in Fig.~\ref{fig:dataset-structure}). This 1000 test-set is used for identifying the accuracy of \emph{matches} of our algorithm during the testing time. To create the \emph{no-match} test-set, we simply remove the 100 apps from 2023 side that were mapped to the 2018 dataset.} 

%% file: sections/4_Methodology.tex
\section{Similarity Search Algorithm}
\label{Sec:Methodology}


Our objective is to identify a 2023 counterpart app for a given 2018 app (a `match') or else to return a `no-match' if such an app does not exist. We follow a multimodal search mechanism to obtain the mappings between the two datasets, and we specifically do not rely on app ID matching in this process. The reason is that metadata-based matching allows us to identify cases of metamorphosis better and to observe whether the developers continued with the same app ID or introduced a new identity.

Our methodology consists of four main steps. 
\textit{
{\bfseries i)} obtain neural embeddings for app icons and app descriptions, and TF-IDF vector representations for app names and developer names, 
{\bfseries ii)} for a given app, retrieve a set of nearest neighbour apps considering each modality, 
{\bfseries iii)} perform modality-wise majority voting to obtain the top-candidate apps for a given app.  
{\bfseries iv)} decide if we could select an app as a `match' from the candidate app list (further discussed in Sec.~\ref{subsec: likelihood_of_continuity}) or else decide it as a `no-match'.} 
We show the end-to-end pipeline of the proposed method in Fig.~\ref{fig:methodology}. 
In the figure, we use the \emph{\textbf{query}} app's embedding representations to retrieve results as \emph{\textbf{search keys}} and the retrieved results (more specifically, retrieved app IDs) as \emph{\textbf{values}}. 
The following subsections explain these steps in detail.

\subsection{Representation of different modalities}
\label{ssec:data_encoding}
As inputs for the search methodology, We use metadata extracted from the Play Store and more specifically, app icon, app description, app name and developer name for both datasets 2018 and 2023. 
We selected these based on prior work and our investigations on the available metadata that are in spotlight for target app users. (Ablation results in selecting the modalities are depicted in Table~1 of the Appendix~D.)
For example, developers use the same name to maintain brand consistency for their apps, compared to the developer email, which is often not publicly available and developers might use different email addresses for different apps to avoid confusion.


\subsubsection{App icon embeddings}
\label{SubSubSec:embeddings}

The app icon serves as the initial point of visual engagement for users, often functioning as a distinctive trademark representing the developer's brand identity. Therefore, a matching pair of apps is likely to have a similar visual `styling' and similar `content' information, emphasising the need to encode both such information.
StyTr$^2$ introduced by \cite{deng2022stytr2} is a popular transformer-based framework with style and content encoders, both producing output embeddings of the shape $(1\times 512)$. Additionally, we experiment with vision transformer content encoder embeddings ~\cite{vit_dosovitskiy2020image} as well as style and content embeddings from VGG19 architecture as proposed in~\cite{karunanayake2020multi}. We identify which style and content embedding combination produces the best results for both `matches' and `no-matches' by evaluating possible combinations against our validation dataset. As presented in Table~\ref{tab:ablations_threshold} we selected the combination of StyTr$^2$ content embeddings and VGG19 style embeddings to represent app icons.





\subsubsection{App description embeddings}
Despite large language generative models being popular, text embedding transformer models, due to their unique encoding capabilities, allow us to obtain rich vector representations for text sequences~\cite{kementchedjhieva2023exploration}. Therefore, we use the MPNet~\cite{song2020mpnet} model to encode the long app descriptions into the vectors, $v_d$ of shape $(1 \times 768)$. MPNet is a language model similar to BERT~\cite{devlin2018bert} and allows to obtain vector representations of text. It considers masked language modelling such as BERT and permuted language modelling such as XLNet~\cite{yang2019xlnet} in a unified view and thus inherits the advantages of both methods. 



\subsubsection{App name and Developer name embeddings}
Since the app and developer names have limited meaningfulness as natural language, as well as they are very short texts, we employ TF-IDF vectorisation to encode app names and developer names instead of using MPNet embeddings. We represent each app and developer name as a vector of size $(1 \times 4,096)$ where the vocabulary contains 4,096 most frequent $1-4\;grams$ when 200,000 app descriptions were used to build the vocabulary.




\subsection{Nearest neighbours of each modality}

Identifying potential apps that underwent ``metamorphosis'' involves taking a query app from the top-10,000 apps in the 2018 dataset and identifying the most similar app (if there is any) among the 1,280,142 odd apps in the 2023 dataset. This requires identifying close apps along each modality presented in Sec.~\ref{ssec:data_encoding} using nearest neighbour search. However, the vanilla version of the nearest neighbour search is highly inefficient here because of the larger queried set, the number of modalities, and the sizes of the embeddings. 


Instead, we use the indexing options provided in the Faiss library~\cite{johnson2019billion} to create an Inverted File Index of the queried embeddings to narrow the search scope. First, we create sharded indexes for small batches of embeddings, and we build the final index on disk by merging sharded indexes into one larger index. We created five such indexes for each modality (represented by an embedding type; app icon - content, app icon - style, app name, app description and developer name), and they are queried for nearest neighbours using the query app's embeddings.

For a single query, we obtain five different nearest neighbour sets corresponding to the five modalities we consider in the descending order of the similarity with the query app's embedding vectors. 


\subsection{Majority voting}
\label{SubSec:majority_voting}

\begin{figure}[ht]
    \centering
    \includegraphics[width=0.48\textwidth]{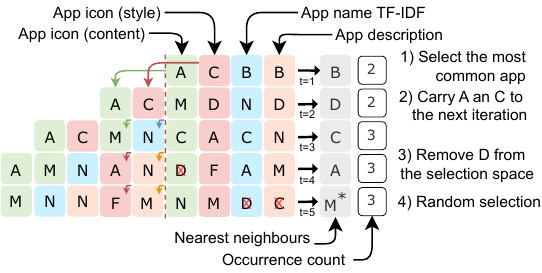}
    \caption{Majority voting}
    \label{fig:majority_voting}
\vspace{-0.3cm}
\end{figure}

Starting of the app similarity search algorithm is a majority voting scheme based on four modalities (we have excluded the `developer name' to be discussed later). Fig.~\ref{fig:majority_voting} emphasises four main steps based on the previously retrieved nearest neighbour lists for four modalities in the descending (vertical) order. In each iteration, we select the app that has the most number of occurrences in each row-wise-position as the most common result. In the figure, each of these selections are visualised in the grey colour box. 

\begin{table*}[h]
  \footnotesize
  \caption{\centering Harmonic mean on the validation set for different embedding combinations and voting thresholds}
  \label{tab:ablations_threshold}
  \centering
  \begin{tabular}{lrrrrrrrrrrrrrrr}
    \cline{2-16}
    & \multicolumn{2}{c}{$\alpha$ = 1} & \multicolumn{2}{c}{$\alpha$ = 2} & \multicolumn{2}{c}{\textbf{$\alpha$ = 3}} & \multicolumn{2}{c}{$\alpha$ = 4} & \multicolumn{2}{c}{$\alpha$ = 5} & \multicolumn{5}{c}{Harmonic mean} \\
    \cline{2-16}
    &      &  No     &     & No      &      & No      &     &  No     &      & No      &       &&    $\alpha$   &       &  \\
    \cline{12-16}
    & Match & Match & Match & Match &  Match & Match & Match & Match &  Match & Match &    1   &   2    &3&    4   &    5    \\
    \hline
    \textbf{StyTr$^2_c$+VGG19$_s$} & 88.8&0.0&	88.4&	36.8&	86.0&	88.2&	63.0&	97.0&	41.8&	99.2&	0.0&	52.0&	\textbf{87.1}&	76.4&	58.8\\
    VGG19$_c$+StyTr$^2_s$ & 89.2&	0.0&	88.8&	37.6&	85.4&	88.2&	62.6&	97.2&	45.2&	99.0&	0.0&	52.8&	86.8&	76.2&	62.1\\
    VGG19$_c$+VGG19$_s$&86.2&	0.0&	85.8&	28.8&	83.6&	86.6&	64.4&	96.2&	49.4&	98.2&	0.0&	43.1&	85.1&	77.2&	65.7\\
    StyTr$^2_s$+ViT & 90.0&	0.0&89.4&	35.4&	86.2&	86.0&	65.8&	96.4&	44.4&	98.8&	0.0&	50.7&86.1&78.2&61.2\\
    ViT$_c$ +VGG19$_s$&88.8&	0.0&	88.4&	32.2&	86.2&	85.6&	70.2&	95.6&	49.8&	98.2&	0.0&	47.2&	85.9&	81.0&	66.1\\
    StyTr$^2_c$ + StyTr$^2_s$&85.0&	0.0&	84.2&	35.4&	80.6&	86.0&	54.0&	96.4&	40.2&	98.8&	0.0&	49.8&	83.2&	69.2&	57.1\\
    \hline
  \end{tabular}
\end{table*}


As shown in Fig.~\ref{fig:majority_voting}, in step 1, app B is selected as the most common occurrence since it appears as the top choice in two separate modalities (occurrence count = 2). Then the remaining apps (A and C) from that position are carried forward to the next iteration. This ensures all potentially similar apps are considered in subsequent row-wise-positions to be iteratively repeated. As shown in step 3, we exclude app D from the selection space since it has already been selected in the second position. When multiple apps appear with the same frequency, we make a random selection. Step 4 indicates this by randomly selecting app M, while both M and N have equal number of occurrences. \emph{The final output is the list of nearest neighbours (vertical list in grey colour) and their corresponding occurrence count for a given query app}.




Note that we handle developer names in a slightly different way as step number 5 that is not shown in Fig.~\ref{fig:majority_voting}. Since some developers publish more than one app, if we use the developer name as a modality in the neighbour search, the result with the highest occurrence count may not be the desired matched result. 
For example, if we have the same developer for app X and app Y, then there is no straightforward way to mention X and Y both in the same position in the app developer neighbour list. 
Therefore, for each row-wise-position, we check the app with the most occurrence count (e.g., app B in step 1) is developed by the same developer of that row-wise-position's app-developer modality result. If so, we consider it as a modality match and add one more to the occurrence count (e.g., app Z in app-developer modality position is the same developer as app B, then we add 1 to the occurrence count in step 1).
We only change the occurrence count list (if applicable) in this step.



\subsection{Match or No-match?}
\label{subsec: likelihood_of_continuity}

Upon retrieving the list of nearest neighbour apps for a given query app, we identify a \emph{matching} app as the first neighbour app in the list's descending order that agrees with following two criteria. 

\begin{itemize}[]
    \item We define a hyper-parameter $\alpha$ such that the number of modalities that contributed for an app to be selected as a nearest neighbour (i.e., the occurrence count) should be equal or greater than $\alpha$. We call this hyper-parameter as `occurrence count threshold'.
    \item We calculate the change of star rating count, representing the number of users who have rated the app, between the neighbour app and the query app. It should be positive. Our intuition here is that a matching app should have grown in rating counts over the past five years. 
\end{itemize}

However, if all the apps in the nearest neighbour search list show a negative star rating count change with respect to the query app, then we only consider the first criteria. But, if no app satisfies the first criteria, we denote it as query app \emph{not having a match}. The threshold value $\alpha$ is selected based on an ablation study we conducted, and we discuss this in Sec.~\ref{Sec:Results}. It is also worth noting here that having a \emph{no-match} result does not mean that we exclude that particular app from metamorphosis analysis. We further utilise them to identify and analyse interesting transitions of apps in Sec.~\ref{Sec:Analysis}.

%% file: sections/5_Results.tex
\section{Performance Analysis of App Similarity Search}
\label{Sec:Results}

In this section, we evaluate the similarity search algorithm based on the validation and test sets described in Sec.~\ref{Sec:Datasets}. To summarise, when we are using the validation-set for identifying \emph{`matches'}, then we have 500 query apps from 2018 to be matched to a counterpart app among 5000 apps in 2023 dataset. When we are identifying \emph{`no-matches'}, we query the same 500 apps from 2018 to be \emph{`not-selected'} among the 4500 in 2023 dataset because the matching 500 apps of 2023 are removed in this setting. Test-sets are used to observe the performance after the hyper-parameters are selected and they work in a similar setting too; 100 apps in 2018 to match among 1000 total in 2023 for `matches' and the same 100 apps among 900 in 2023 for `no-matches'. Next we will discuss the hyper-parameters we validate our model with. We make two hyperparameter choices based on the performance on the validation set.

\begin{itemize}[]
    \item We explore six different combinations of style and content embeddings.
    \item We explore five different occurrence count thresholds ($\alpha$ values) from 1 to 5. 
\end{itemize}

We evaluate a occurrence count threshold, $\alpha$ for each embedding combination and report the harmonic mean of the two scenarios. We show the results in Table~\ref{tab:ablations_threshold}. As can be seen, the combination of StyTr$^2$ content embeddings and VGG19 style embeddings produce the best performance at the voting threshold $\alpha=3$. We used this hyperparameter combination to obtain the main results of the paper.

We also observe that the accuracy of the no-match scenario increases when we increase the thresholds. This is expected because as we increase the thresholds, the number of modalities an app should be included to select as a match increases. On the other hand, the accuracy decreases with the increased thresholds when there is a match. This happens because, as the threshold increases, an app should be included in more modalities in order to be selected as a match, therefore, the conditions become more restrictive. The following subsections describe the performance metrics we use and the performance of the test set.

\subsection{Performance metrics}

We measure the performance of our method on test set and compare with an existing baseline using accuracy, precision@1 and recall@1 matrices. 

\subsubsection{\textbf{Accuracy}} We take the harmonic mean of the accuracies when there is a \emph{match} and when there is \emph{no-match}. We use harmonic mean instead of the arithmetic mean because we need accuracy in both scenarios to contribute equally.

\subsubsection{\textbf{Precision and Recall}}
We define Precision and Recall based on the following definitions. For a match scenario, there can be \textit{i) a correct match (true positive), ii) an incorrect match (false positive)}, and \textit{iii) no result when there is a correct match available (false negative)}. Note that there are no true negative occurrences since we are considering matches and no-matches separately. We define the Precision and Recall for a match scenario as in Eq.~\ref{eq:precision} and Eq.~\ref{eq:recall}.

\begin{small}
    \begin{equation}
    \label{eq:precision}
        Precision = \frac{|True\,Positives|}{|True\,Positives| + |False\,Positives|}
    \end{equation}
    
    \begin{equation}
    \label{eq:recall}
        Recall = \frac{|True\,Positives|}{|True\,Positives|  + |False\,Negatives|}
    \end{equation}    
\end{small}

For non-match scenarios, the algorithm can output either no match (True Positive) or an incorrect match (False Negative). Therefore we define only Recall for the non-match as defined in Eq.~\ref{eq:recall}. Note that Recall and Accuracy are similar for the no-match scenario.

When we report our results as precision@1 and recall@1, we only consider the first matching result for each method when calculating scores.


\subsection{Performance analysis}
\label{SubSec:PerformanceAnalysis}

We compare our method with the method proposed by  Karunanayaka et al.~\cite{karunanayake2020multi}. In~\cite{karunanayake2020multi}, the authors proposed a nearest neighbour search for the weighted sum of each embedding, including the app icon (style and content) and app description. We adapt their method using $\alpha, \beta, \theta, \gamma \, and \, \delta$ \cite{karunanayake2020multi} as the weights for image content embeddings, image style embeddings, developer name embedding, app description embedding and app name embeddings, respectively. In addition to those weights, we use an additional threshold parameter to decide the apps that do not have any matching apps. Then, we choose the optimal values for those weights and the threshold using the Bayesian optimisation \cite{fer2014nog}  on the validation dataset considering the integer values between 1 and 10. We report the results for the best-performed values of $\alpha, \beta, \theta, \gamma, \delta$ and the threshold and results for our method in Table~\ref{tab:comparing_with_naveen}.


\begin{table}[ht!]
  \footnotesize
  \caption{\centering Similarity search performance on the test set}
  \label{tab:comparing_with_naveen}
  \centering
  \begin{tabular}{lrrrc}   
    \cline{2-5}
    &   \multicolumn{3}{c}{Match} & No match \\
    &  A@1 & P@1 & R@1 & R@1 \\
    \hline
    Majority voting (ours) &  \textbf{0.930} & 0.989 & \textbf{0.939} & 0.900\\
    \hline
    $\alpha, \beta, \theta, \gamma, \delta~\cite{karunanayake2020multi}$ & 0.870 & \textbf{1.000} & 0.870 & \textbf{0.980}\\
    \hline
  \end{tabular}
\end{table}

As can be seen from the results, our method provides better results when the app is included within the search space (i.e., with \emph{matches}). Conversely, the method proposed by~\cite{karunanayake2020multi} demonstrates higher accuracy for \emph{no-matches}. 
Nonetheless, As shown in Appendix~B, \cite{karunanayake2020multi} becomes infeasible to use when the queried dataset and the number of search keys are larger, given that it uses a brute force approach, comparing the search key with embeddings of all apps in the queried data multiple times for different types of embeddings.

\begin{figure*}[t]
    \centering
    \includegraphics[width=0.99\textwidth]{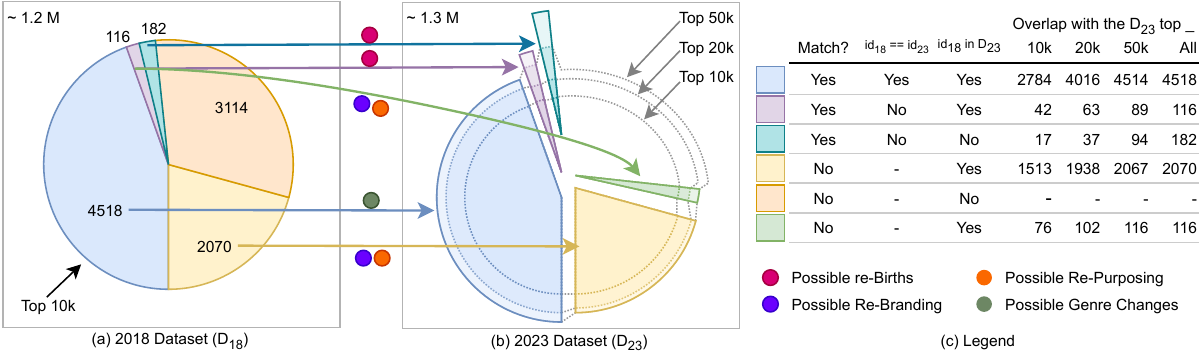}
    \caption{Identifiable mappings and regions of interests that are obtainable from our similarity matching algorithm. Area under each pie segment is indicative of the proportion of apps belonging to each category as described in the legend. Note: top 10k, 20k and 50k dashed lines are not to the scale, but all pie charts and the proportions they represent between top10k, 20k and 50k are true-to-the scale for better visualisation.}
    \label{fig:mappings_from_our_method}
\end{figure*}

\subsection{Success score}
\label{SubSec:success_score}

We introduce \emph{`success score' (SS)} as a parameter that we can use to evaluate the performance of each app progressing from 2018 to 2023. In high-level, we expect a positive SS to indicate an app that has been performing well such that there is no risk of that app going obsolete and a negative SS to indicate an app not performing well.  


The success of an Android app could intuitively be associated with the number of downloads and the number of ratings it received from the app users, both ultimately impacting the direct and in-direct revenue incoming to the developers. As Google Play Store only provides end-users with the cumulative download counts and the rating counts as app metadata, it does not reflect current installations (active users) or indicate whether an app is gaining popularity or losing user interest. Therefore, it is relevant to compare the increase or decrease of these numbers across a period of time, as in our study, across five-years. However, an increase in these counts does not necessarily mean an app is performing better due to the natural growth of the Android user base over a period of time: 2.3 billion worldwide devices in 2018~\cite{bengoogle} to approximately 3.6 billion in 2023~\cite{ashgoogle,avada, statista, newzoo}. Therefore to distinguish an over-performing app and an under-performing app, we need to adjust this natural growth rate which is likely to have been incorporated to any app in the app market.

To address this limitation, we propose a metric, referred to as the success score, which offsets the average Android user base growth against the average growth in download and user rating counts for a given app. We employ the widely used~\cite{peakcagr} Compound Annual Growth Rate (CAGR) to analyse the growth or decline of the three key metrics: the number of app downloads, the number of user ratings, and the number of Android device users. Since our dataset spans five-year intervals, CAGR smooths out short-term volatility for a more consistent measure of growth and is ideal for capturing long-term changes by presenting an average annualised growth rate unlike the annual growth
rate (AGR) which focuses on year-to-year variations~\cite{jsoncagr, markcagr}.

Considering this factor, we define the \textbf{Success Score (SS) for an app based on the CAGR of downloads and ratings offset by the natural growth in the Android ecosystem over a time span of $t$ years}. As specified in Eq.~\ref{eq_cagr}, when we calculate $CAGR_{\#downloads}$, the term $n_{initial}$ is the download count of 2018 app, $n_{final}$ is the download count of counterpart 2023 app and $t=5$ years. $CAGR_{\#ratings}$ is similarly calculated based on the star rating count and $CAGR_{\#Andro}$ is calculated based on the increase of android devices over the five years, 2.3B to 3.6B.


\begin{equation}
    CAGR = (\frac{n_{final}}{n_{initial}})^{\frac{1}{t} - 1}
    \vspace{-0.1cm}
    \label{eq_cagr}
\end{equation}

\begin{equation}
    SS = \frac{1}{2}CAGR_{\#downloads} + \frac{1}{2}CAGR_{\#ratings} - CAGR_{\#Andro}
    \label{equation_ss}
\end{equation}    

According to the Eq.~\ref{equation_ss}, SS value of $x$ indicates that the average number of downloads and user ratings have increased $x\%$ better than the average growth of android eco-system growth per year. Moving forwards, we discuss SS as a percentage (e.g. SS=0.145 is interpreted as 14.5\%) for convenience.

%% file: sections/6_ResultsAnalysis.tex
\section{Characterisation of App Metamorphosis}
\label{Sec:Analysis}

In this section, we provide a characterisation of different types of metamorphosis categories listed in Fig.~\ref{fig:re-emergence tree diagram}. We explain how to identify each category using our search methodology, analysis and insights on the causes and effects of metamorphosis, followed by interesting examples.




First, we use our search algorithm on the top 10,000 apps from the 2018 dataset (Fig.~\ref{fig:mappings_from_our_method}a) to identify their corresponding apps among the 1.3 million apps in the 2023 dataset (Fig.~\ref{fig:mappings_from_our_method}b). The legend in Fig.~\ref{fig:mappings_from_our_method}c indicates how we identify the significant metamorphism regions of interest in D18 to D23 mappings. First, we check if our algorithm gives a \emph{match} or a \emph{no-match}. If it is a \emph{match}, we observe if the app ID in 2018 is identical to the app result we get in 2023. If they are not identical (i.e., 2018 app ID $\neq$ 2023 app ID), then we check whether the 2018 original app ID exists in the 2023 dataset; if it exists, we analyse them further. Lastly, for any of the 2018 top 10k apps that our algorithm gives \emph{no-match}, we check again if those 2018 app IDs exist in the 2023 dataset and analyse further.


\subsection{Summary of mappings}
\label{sec:summary_of_mappings}

As shown in Fig.~\ref{fig:mappings_from_our_method}, a total of 4,518 apps get a \emph{match} in D23 having the same app ID between the two datasets (blue arrow). Nonetheless, only 61.6\% of them are still among the Top 10k of D23. The rest were mostly in between the top 10k and 50k.  
We observe an average success score (SS) of 10.01\% for these 4,518 apps, and this score is more influenced by download count ($\sim$20\% more contribution) rather than user rating. These apps are likely not to have gone through drastic transformations as our algorithm has already returned a \emph{match}, and the app IDs are the same between the two datasets. We use their SS numbers as the baseline to compare with the SS numbers of different metamorphosis categories. 

We observe 298 apps that our algorithm \emph{matched}, without the same app ID among the matching pairs (teal and purple arrows). They suggest that the original app continued to 2023 with a different identity, which we identify as a potential \emph{re-birth} that is discussed later in Sec.~\ref{subsec_reborn_apps}. Out of those 298, 116 had the same app ID linked to other apps in the 2023 dataset (green arrow) that our algorithm did not find as similar. Furthermore, there are 2,070 other examples in the 2018 dataset initially returned again as \emph{no-matches}, but the same app ID exists in the 2023 dataset (yellow arrow). Both of these scenarios suggest that the developer has decided to \emph{re-brand} or \emph{re-purpose} those apps, resulting in major changes to the app description, name and app icon; hence why our methods did not match them. We discuss this further in Sec.~\ref{sec:rebrand} and Sec.~\ref{subsec_repurposed_apps}.


We also discuss other interesting adaptations by the app developers, such as changes in the genre, content-rating, revenue-model and developer-transitions. Furthermore, developers might opt to keep multiple versions of the same app or could develop apps targeted at particular user demographics. We discuss them in detail in Sec.~\ref{subsec_genre_change} to Sec.~\ref{subsec_versions}. 

Finally, there are 3,114 apps, for which our method gives \emph{no matches}, and the same app IDs are no longer present in D23 (orange arrow), indicating either the developers have discontinued those apps or they have been removed by Google (e.g. due to spamming~\cite{seneviratne2015early}). When an app is discontinued, it is likely caused by the applications of such apps no longer being required or being replaced by progressive versions. Majority of these discontinuations were from tech giants such as \emph{Google, Samsung, HTC} or from popular game developers such as \emph{Gameloft}, and \emph{Electronic Arts}.

\subsection{Re-born apps}
\label{subsec_reborn_apps}

Re-born apps are characterised by their initial discontinuation, followed by a reappearance where either the original developer or a new one reintroduces the same app concept. \\ \vspace{-3mm}

\noindent{\textbf{Identification:} We select \emph{matched} apps from the similarity search where the 2018 app ID is different from the 2023 app ID. We found a total of 298 belonging to this condition ({\bf cf.} Fig.~\ref{fig:mappings_from_our_method}: 116 in purple and 182 in teal).}
We further filter them by selecting the apps with a release day in D23, which is later than the last update day in D18, and obtain 88 potential apps for re-birth. We manually validated these 88 apps and found that 74 of them (84.1\%) are indeed actual re-births, indicating our identification method works. \\ \vspace{-3mm}

\noindent{{\bf Analysis - } Because of the app ID change, any market presence the 2018 app gained is discontinued and the new app has to start afresh. As a result, the average success score (SS) for an app in this category is -14.81\%, and they struggle to be within the top 10k category in D23 as further illustrated in Fig.~\ref{fig:mappings_from_our_method} with only 20.4\% of the apps represented by purple and teal colours are in the top 10k of 2023. We note the following observations while analysing the re-born apps.} \\ \vspace{-3mm}

\begin{figure*}[t]
    \centering
    \includegraphics[width=0.99\textwidth]{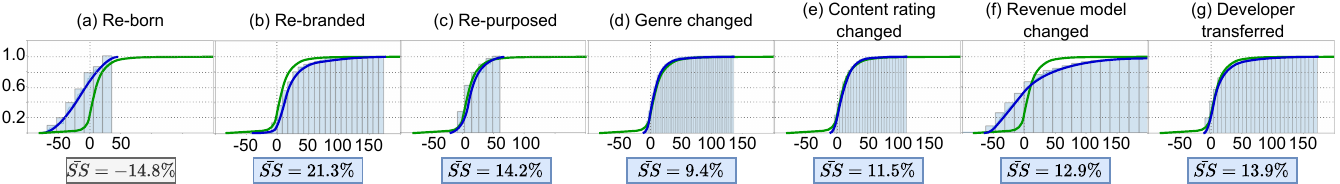}
    \caption{CDF plots for the selected metamorphosis categories. X axis represents the success score (SS \%). $\bar{SS}$ represents the average SS for all apps undergoing the transition per sub-figure. Legend: Blue: smoothed CDF plot. Green: Baseline CDF for the SS of 4,518 app pairs where our model gives a \emph{match}.  }
    \label{fig:cdf_figures}
\end{figure*}
\begin{figure*}[ht]
    \centering
    \includegraphics[width=0.99\textwidth]{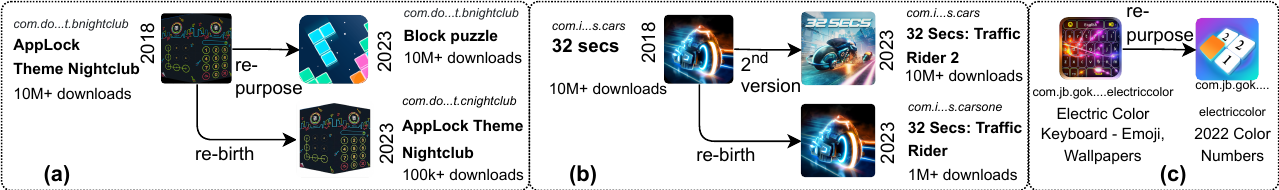}
    \caption{(a) a special example where an app re-purposed and a re-birth occurred using a different app ID. (b) an example where an app changed to a different version, however the original app re-born again using a different app ID (c) a generic example of an app re-purposed }
    \label{fig:fig_repurposed_apps}
\end{figure*}

\begin{itemize}[]
    \item In Fig.~\ref{fig:cdf_figures}(a), we plot the cumulative distribution function (CDF) of SS scores of re-born apps. It further shows that the success of re-born apps is inferior compared to the baseline SS-CDF of nearly 5,000 matched apps between the datasets, which remain in the top 10k ({\bf cf.} Sec.~\ref{sec:summary_of_mappings}). Here, we also observe that nearly $\sim70\%$ of apps in the re-born category have a negative SS compared to $\sim32\%$ in the baseline CDF.
    
    
    \item Despite the negative average SS in this category, 30.13\% of apps such as, \emph{Bowmasters, MONOPOLY, VLC for Android} have managed to recover (SS>0) and even build upon their user base and ratings within or less than five years time span. 
    
    \item Games are often ($\sim27\%$) re-introducing the progressive versions as re-births. An already established user base would actively follow newer versions, therefore contributing to pushing the SS towards a positive value. This allows the developers to discontinue the old version after the transition period to reduce the maintenance cost of multiple apps. Changing the original app to a newer version while re-birthing the former using a new app ID is very rare. But we observed one such example as shown in Fig.~\ref{fig:fig_repurposed_apps}(b).

    \item We identify that sometimes developers change their original app into a totally new type of app using the same app ID (i.e., re-purposed - discussed more in Sub.Sec.~\ref{subsec_repurposed_apps}). However, they do not want to discontinue their original app to keep the revenue stream. They get a new app ID and introduce the original app again, which is a re-birth. For example, \emph{AppLock@DoMobile} changed their \emph{AppLock Theme Nightclub} app to a game called \emph{Block puzzle} and re-introduced the \emph{AppLock Theme Nightclub} under a different app ID ({\bf cf.} Fig.~\ref{fig:fig_repurposed_apps}(a)). 
    

    \item 29.5\% of manually verified re-births happen with developer name changes. Even if the app name, description and visual data allow us to verify them as the same app being re-introduced, a malicious actor could easily employ the same strategy and introduce counterfeit apps as re-births when one app gets discontinued. This puts even a tech-savvy user in a vulnerable position as Google Play Store does not provide the history of an app developer. We further discuss this in Sec.~\ref{sec:sec_security}. 
\end{itemize}



\begin{figure}[ht]
    \centering
    \includegraphics[width=0.48\textwidth]{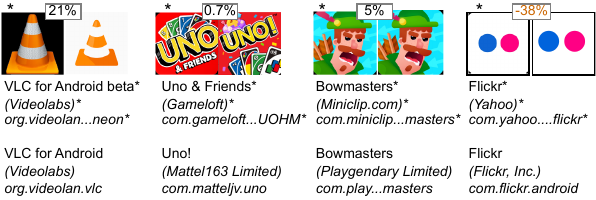}
    \caption{Examples for re-birth. Mentioned in italics are the developer name and app ID. (*: indicates 2018 version.) The success score for the transition is numbered inside the text box.}
    \label{fig:rebirth_examples}
    \vspace{-0.3cm}
\end{figure}

\noindent{\bf{Examples}} - We have identified notable apps while observing this category, including \emph{Uno}~\cite{uno} and \emph{Flickr}~\cite{flickr} ({\bf cf.} Fig.~\ref{fig:rebirth_examples}). In both of those examples, the original app lost the existing user base as the app ID was discontinued. However, different developers re-introduced very similar apps without a major change in the functionality. In 2019, \emph{SmugMug} acquired Flickr from the former owner \emph{Yahoo} and reintroduced it as a new app with significant changes. However, the SS of the app is -38.8\%, indicating that it experienced a significant loss of its user base during this transition. Conversely, the $Uno$ app has a positive SS of 0.7\%. Despite a decrease in its rating count, as reflected by a CAGR$_{ratings}$ of -0.3\%, games like $Uno$ naturally has a high demand, evidenced by CAGR$_{downloads}$ of 20.2\%.

\subsection{Re-branded apps} 
\label{sec:rebrand}

Some apps in Google Play can stagnate after a while because they no longer attract new users, stalling the growth of app-related revenue. In such settings, a developer may opt to change an app's outlook drastically, fine-tune some degree with features or perhaps may opt to transfer/sell the app to a new developer, again resulting in drastic changes to the app according to new ownership. We identify this change as app re-branding. \\ \vspace{-3mm}

\noindent{\textbf{Identification}: We select \emph{non-matched} apps from similarity search where the 2018 app ID still exists in 2023 dataset ({\bf cf.} Fig.~\ref{fig:mappings_from_our_method}: 116 (green) and 2,070 (yellow)). A re-branded app could potentially have changed in the outlook (app icon, visual style and colour schemes, app name) but should have a similar context in the app description. (A drastic change in app description indicates that the app may have been re-purposed rather than re-branded). Therefore, we further filter previous results by selecting the apps with app name and app icon content embeddings having a smaller cosine similarity (< 0.7; i.e., these features are likely changed) and app description cosine similarity being higher (>0.4; i.e., still the descriptions remain relatively unchanged) between 2018 and 2023 counterparts.} \\ \vspace{-3mm}


\noindent{\bf{Analysis:}} From previous criteria, we retrieved 322 apps that potentially underwent re-branding. Since we know the app ID is identical for all of them across the 2018 and 2023 datasets, there should not be false retrievals. Yet, we manually evaluated all of them for characterisation. 66.8\% (215 apps total) of them portrayed an identifiable reason for re-branding; for example, the app name/app description mentioned the particular changes or the developer was changed along with modifications. The remaining apps had drastic changes in the app name and description, but we were unable to determine a specific root cause. Furthermore, out of the verified 215, 38.6\% of the re-branding occurred within the same developer. \\ \vspace{-3mm}

\begin{figure}[ht]
    \centering
    \includegraphics[width=0.48\textwidth]{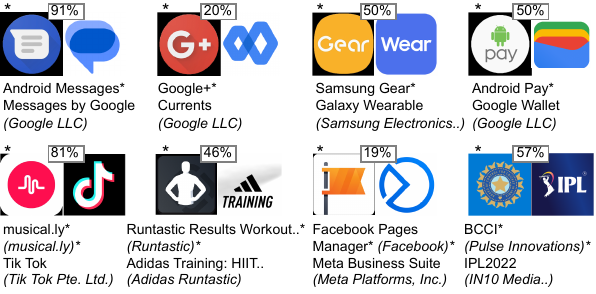}
    \caption{Examples for re-branding. (*: indicates 2018 version.) The success score for the transition is numbered inside the text box. }
    \label{fig:rebranded_examples}
\end{figure}

\noindent{\bf{Examples}} - Our SS calculation for re-brand apps is 21.3\% which is notably 11.3\% higher than the baseline average SS for $\sim$5,000 matched apps. This is further emphasised in the Fig.~\ref{fig:cdf_figures}(b) establishing that re-branding has a better chance of success among all other categories of metamorphosis. 
This positive SS is anticipated as re-branded apps retain their download and rating counts during the transition and re-branding potentially brings new user engagements. Notably, \emph{Mi Drop} to \emph{ShareMe: File sharing} by Xiaomi Inc, \emph{Android Messages} to \emph{Messages by Google} and \emph{Google Duo} to \emph{Google Meet}, each achieving a SS over 88\%. Google LLC claims that the transition of Google Duo to Google Meet is to create a unified platform to provide users with an enhanced experience by adding new Google Meet features to the Duo app. Apps such as \emph{Musical.ly} were re-branded by merging with an existing app like \emph{TikTok} to gain international recognition. Another notable example is Google introducing \emph{Google Currents} as a replacement for \emph{Google+} which has a SS of 10.98\%. According to G Suite, this is  due to low user engagement and security concerns. Even though Google Current is designed for organisations, the app's main purpose remains the same. This type of overhaul creates new discussion topics among the community to attract more users and also enhance the core functionality the original app was based on. This is quite prominent among tech giants such as \emph{Google, Microsoft} where there is considerable media attention to anything novel coming out of them. As shown in Fig.~\ref{fig:rebranded_examples}, other notable examples include \emph{Android Pay} to \emph{Google Wallet}, \emph{Google Fit: Fitness Tracking} to \emph{Google Fit: Activity Tracking} and \emph{Android Messages} to \emph{Messages by Google}. To summarise, re-branding apps tend to show higher success in terms of SS compared to typical apps in the app market, indicating that strategic re-branding can significantly enhance an app's market performance.

\subsection{Re-purposed apps} 
\label{subsec_repurposed_apps}

Contrary to what is discussed before, some apps may experience quite a significant loss of interest over time and may no longer provide useful services to the user base. For example, rapid developments in the Android operating system could cause many popular apps such as camera, file management, battery optimisation and other utility apps to become obsolete quickly. The developers may want to relaunch them by overhauling the original functionality; we call them "re-purposed" apps. The basic intuition behind this incentive is that the original app has a significant user base and contains good reviews, and the developers need to build on top of that. There could also be instances where a developer needs to transfer the app to a blooming app category, away from the existing competition but intending to maintain the previous user base. Overall, reasons to re-purpose an app compared to the original app concept could be summarised as follows;

\begin{itemize}[]
    \item The developers perform drastic changes in core functionality; therefore, the original app is re-purposed, and the original concept is now discontinued. This can also be considered as an extreme end of app re-branding.
    \item The developers re-purposed the original app. However, they intend to retain their original concept and reintroduce it as a re-birth ( {\bf cf.} Sec.~\ref{subsec_reborn_apps}). 
\end{itemize}

\noindent{\textbf{Identification}: How we identify re-purposed apps is the same as identifying re-branded apps. However, we only retrieve instances where the app description's cosine similarity is between 0.2 and 0.4. The reason is that a re-purposed app drastically changes functionality, and the app description embeddings should deviate significantly from one another. Similarities of less than 0.2 were observed to be noisy, such as descriptions in two different languages, and as a result, we discarded them from the evaluation. The selection criteria is summarised in Appendix~E.} \\ \vspace{-3mm}

\noindent{\bf{Analysis:}} The average SS for this category is 14.3\%, which is lesser than the SS of re-branded apps yet higher than the baseline as shown in Fig.~\ref{fig:cdf_figures}(c). This could be due to the nature of re-purposing, which involves significantly changing the app's functionality or target audience, which can negatively affect its existing user base. In contrast, re-branding maintains its core functionality and changes the brand elements. Overall, we identified 33 re-purposed examples, and we manually verified all of them. We identified 24.2\% of them as ideal examples. We couldn't conclusively verify the rest of the apps, which will likely be at the intersection of re-branding and re-purposing. \\ \vspace{-3mm}

\noindent{\bf{Examples}} - We show some examples of re-purposing in Fig.~\ref{fig:fig_repurposed_apps}(a) and Fig.~\ref{fig:fig_repurposed_apps}(c). The \emph{AppLock Theme Nightclub} app was re-purposed to a \emph{Block Puzzle} game, while \emph{Electric Color Keyboard} was re-purposed in to \emph{2022 Color Numbers} apps. Both of these examples highlight how developers leveraged their existing user base to introduce new app concepts. Some other notable examples include; \emph{Fingerprint Lock Screen - Prank} (10M+ downloads) re-purposed to \emph{GPS Navigation - Route Finder} (10M+ downloads) (app ID: \texttt{com.galaxyapps.lock}). An incentive from developers like this is undesirable for app users as they did not download the original app for a navigational purpose.




\subsection{Genre changes}
\label{subsec_genre_change}

Genre changes, also known as category changes, represent the instances where the high-level category (e.g., Educational, Simulation, Casual) is changed in an app. We obtain this information from the app metadata hosted in Google Play Store for each app.  \\ \vspace{-3mm}

\begin{figure}[ht]
    \centering
    \includegraphics[width=0.45\textwidth]{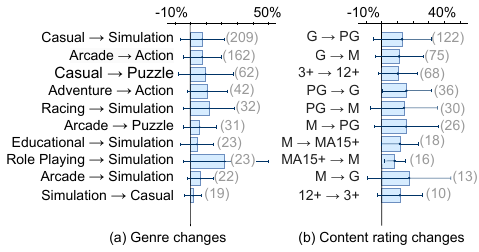}
    \caption{Diagram of 10 most common genre (a) and content rating (b) changes with respect to their average success score. Standard deviation of the SS for each transition is also shown using error-bars. The number enclosed in parenthesis represents how many apps underwent the transition.}
    \vspace{-0.3cm}
    \label{fig:genre_and_content_rating_changes}
    
\end{figure}

\noindent{\bf{Identification}} - We discover a total of 1,111 alterations in the app categories among the 4,518 \emph{match} and \emph{2018 app ID = 2023 app ID} instances identified by our algorithm. 
We present the most commonly observed category changes in Fig.~\ref{fig:genre_and_content_rating_changes}(a). \\ \vspace{-3mm}

\noindent{\bf{Analysis and Examples}} - The average SS for apps that have undergone a genre change is 9.44\% as shown in Fig.~\ref{fig:cdf_figures} (d). Therefore, despite the app undergoing genre changes, it remains as competitive as an average top 10k app on average. As shown in Fig.~\ref{fig:genre_and_content_rating_changes}(a), the most frequent genre transition involves casual apps shifting to simulations. Even though 209 apps have made this transition, their SS is lower than the average at 7.74\%. However, 23 apps that became simulation from being role playing, have achieved an average SS of 22.14\%. 


The motivation behind altering the app category/genre by developers can stem from several plausible reasons, including i) the aim to better align the app with its features, functionality, or target audience; ii) adjusting to app market changes (e.g., from "Candy Crush Saga" categorized as "Casual" to "Puzzle"); and perhaps iii) move the app by modifying the app category to go into a less competitive app category or to avoid scrutiny (e.g., "CVS Pharmacy" categorised as "Health \& Fitness" to "Shopping"). There has been some evidence for this behaviour in the past~\cite{surian2017app}. 

\subsection{Content rating changes}
\label{content_rating_changes}

Content rating, also known as age rating in Android, is self-reported by the developer. This information is also available in the app metadata in the Play Store. In Australia, where we conducted the data collection, games follow a content maturity rating issued by the local regulatory body (i.e., G, PG, M, MA15+, and R18+), and other apps follow the International Age Rating Coalition (IARC) rating (i.e., Ages 3+, 7+, 12+, 16+, and 18+). Over the past five years, we observe a noticeable shift in content ratings of apps. \\ \vspace{-3mm}

\noindent{\textbf{Identification} - We were able to identify 504 apps with different content ratings among the 4,518 \emph{match} and \emph{2018 app ID = 2023 app ID} instances. We present the most commonly observed changes in Fig.~\ref{fig:genre_and_content_rating_changes}(b).} \\ \vspace{-3mm}

\noindent{\textbf{Analysis and Examples} -
Overall SS for the 504 apps that have changed the content rating is $11.54\%$ as shown in Fig.~\ref{fig:cdf_figures} (e). Within this, we identified 11 apps that has been re-born with changes in content ranting, for example \emph{Uno} shifted its content rating from G to PG after being taken over by a new developer. Further, 17 apps became games and 12 games transitioned into generic apps. However, upon manual analysis, we found that those 17 apps that became games were games in the first place, and the developer content rating was not correctly reflecting that. As an example, the game `Tanktastic 3D Tanks' in 2018 was rated for 3+ and in 2023 (giving a false illusion that this could be a generic app instead of a game), the rating changed to PG (Parental Guidance) and a better match for the overall theme of tank battles.}

We observe a similar setting among game to app transitions as well. Furthermore, as shown in Fig.~\ref{fig:genre_and_content_rating_changes}(b), the most common change happening among game apps is 122 of them increasing their rating from G (General) to PG. A popular game of `Mafia City' changed from PG to M due to violence, strong language and blood-related themes in the game.

We attribute the changes discussed above as better alignment with Google Play Store rating guidelines and more developers being aware of any previous mistake they did. Further restricting a content rating category than the original one is not necessarily adverse. However, whether the existing users were duly notified is questionable. On the other hand, loosening the content rating category could attract more popularity among a wide range of users if they comply with content-rating regulations. Among the 26 games that changed the rating from M to PG, we identified 14 of them directly contained Gun violence related app icons, which is questionable. Since identifying content compliance malpractices is not our primary concern, we open this topic for future research. 




\subsection{Changes in revenue model}
\label{appendix_revenue}

In this context, we consider only the direct revenue model where an app is either "free" or requires payment for installation (hereafter referred to as "paid apps").  

\noindent{\bf{Identification}} - This information is extractable directly via app metadata, and we can utilise our similarity search algorithm to obtain matching pairs and identify the instances of paid apps becoming free. In 2018, there were 38 paid apps, however 9 of them discontinued by 2023. 

\noindent{\bf{Examples}} - Among the results we obtained, we did not observe any examples within the top 10,000 apps of the 2018 dataset that changed from a paid app to a free app. It is understandable that popular paid apps may not consider moving to a free model, as they would not want to lose their existing users who bring direct revenue.
Nonetheless, upon further investigation using the entire 2018 dataset, we could identify 649 paid apps in 2018 that have become free apps with an average SS of $\sim13\%$ (cf. Fig.~\ref{fig:cdf_figures}-f). A noteworthy example is \emph{TextGrabber – image to text: OCR} where originally it was a paid app (\$6.99) but was later changed to a free app, possibly due to many similar tools being available elsewhere for free and the developer deciding freemium might be a better business model than simply offering a paid app. Moreover, \emph{Farm Simulator 16} transitioned to a free app and its download count jumped from less than 1M in 2018 to 65M, securing a top spot in 2023.

\subsection{Transferred apps}
\label{subsec_transferred_apps}

There is an option in Google Play Store for an app to be transferred from one developer to another developer~\cite{devtransferandroid}. Such transfers may occur when developers adopt new business models centered around rapidly growing mobile app concepts (e.g., spin-offs) or when their businesses are acquired by different companies. \\ \vspace{-3mm}

\noindent{\bf{Identification}} - To identify these types of apps, we employ our search method to find \emph{matches} where 2018 app ID = 2023 app ID. From these selections, we extract the ones with a cosine similarity of the developer's name, email, and website below a threshold of $0.5$. In this way, we identified a total of 104 potential app transfers with an average SS of $\sim14\%$ (cf. Fig.~\ref{fig:cdf_figures}-g). \emph{Note: This is not to be confused with `Re-branding' where our algorithm does not give a `match' due to the considerable change in app's appearance. Here, it is simply handed over without major changes.} \\ \vspace{-3mm}

\noindent{\bf{Examples}} - \emph{Subway Surfers}, the first game to cross $1$ billion downloads on the Google Play Store, was initially co-developed by both \emph{Kiloo} and \emph{SYBO Games}. However, \emph{Kiloo} has withdrawn from the development and the game is now solely owned by \emph{SYBO Games}~\cite{subway}. Another example is shown in Fig.~\ref{fig:re-emergence tree diagram} where \emph{Flip Master} game is transferred from popular developer \emph{Miniclip} to \emph{MotionVolt Games}.
Overall, app transfers between developers can bring fresh perspectives, resources, and opportunities for growth, ensuring the continued development, support, and success of such mobile apps.

\subsection{Changes in target demography}
\label{subsec_deomograph}
\begin{figure}[t]
    \centering
    \includegraphics[width=0.35\textwidth]{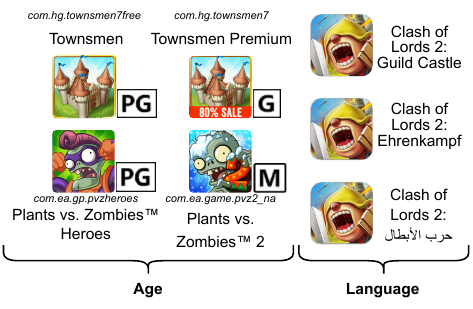}
    \caption{Some examples of apps where the target demography changed based on age and language. }
    \label{fig:fig_demography_change}
\end{figure}

Some developers may opt to publish different variations of the same base app for different target audiences. In previous subsections, we retrieved only the top result of the 5 nearest neighbours by our similarity search algorithm. Observing the results for the remaining nearest neighbours with the condition developer name 2018 $\approx$ developer name 2023, we observe apps belonging to this category.  \\ \vspace{-3mm}

\noindent{\textbf{Age}} - There are 84 apps that are targeting users of different age-demographics with the same concept.  Examples include apps such as \emph{Plants vs Zombies™ 2} (Mature) and \emph{Plants vs. Zombies™ Heroes} (Parental Guidance), \emph{Vector 2} (General) and \emph{Vector 2 Premium} (Parental Guidance), \emph{Silly Sausage in Meat Land} (Mature) and \emph{Silly Sausage: Doggy Dessert} (General). \\ \vspace{-3mm}

\noindent{\textbf{Language and geographical region}} - We noticed 21 apps with multiple languages or target localities. For example, \emph{Clash of Lords 2} has multiple versions for different languages as shown in Fig.~\ref{fig:fig_demography_change} and \emph{TikTok} has two different versions; \textit{com.ss.android.ugc.trill} and \textit{com.zhiliaoapp.musically}. The former version primarily targets East Asian and Southeast Asian countries, while the latter is intended for other countries. \\ \vspace{-3mm}

\noindent{\textbf{Multiple revenue models}} - Exapanding the analysis from previous subsection about changes to the original revenue model, we observed 88 apps that exhibit multiple price variations, with some being offered as free apps and others requiring a payment. A notable example is Need for Speed$^{TM}$ Most Wanted with a paid app; \emph{com.ea.games.nfs13\_na} and a free app; \emph{com.ea.games.nfs13\_row} (only available for some mobile devices).

\subsection{Progressive versions of apps}
\label{subsec_versions}

\begin{figure}[t]
    \centering
    \includegraphics[width=0.48\textwidth]{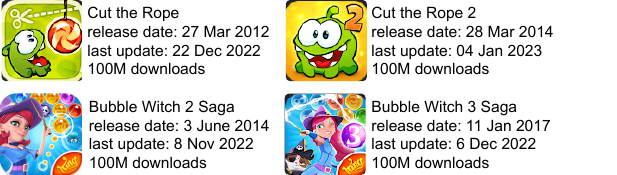}
    \caption{Progressive version examples for two apps, \emph{Cut the Rope} and \emph{Bubble Witch Saga}.}
    \label{fig:version_changes}
\end{figure}

The decision to release newer versions of an app gives opportunity for the developers to enhance features, address bugs, and meet evolving security and technological requirements based on user feedback and suggestions. We discussed in Sub. Sec. ~\ref{subsec_reborn_apps} that most of the games abandon previous versions after some time when the user base has shifted to the newly released version. However, some developers may opt to maintain retain the previous versions such that they can target different niche markets, address specific platforms or hardware compatibility issues, or experiment with novel concepts. Later, based on user feedback and market response, the developer has the flexibility to decide whether to continue maintaining both versions or discontinue one of them. We observe 380 examples of 2018 top 10,000 dataset now has multiple versions by observing the nearest neighbours of our similarity search algorithm with the condition of developer name 2018 $\approx$ developer name 2023. We highlight some examples in Fig.~\ref{fig:version_changes}.

\begin{figure}[ht]
    \centering
    \includegraphics[width=0.49\textwidth]{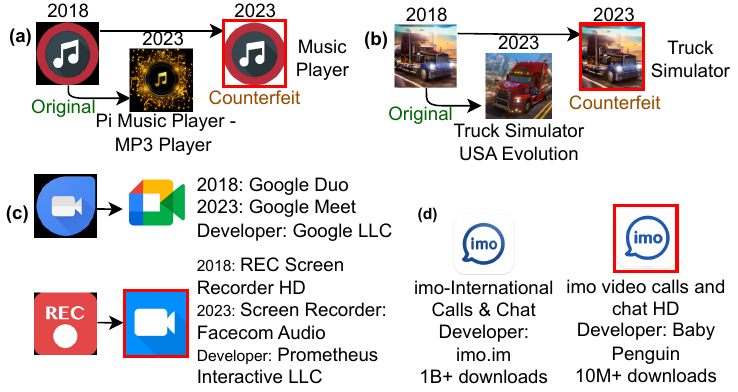}
    \caption{Security Risks of Mobile App Metamorphosis. Outlined in red are observed to be counterfeits/practicing plagiarism.}
    \label{fig:sec_risks}
    \vspace{-0.3cm}
\end{figure}


%% file: sections/7_Discussion.tex
\section{Privacy and Security Implications of Metamorphosis}
\label{sec:sec_security}

Even if the general belief is that the top apps in Google Play Store are trustworthy and are better moderated, our results showed that an app's visual appearance, textual description, developer name, or all of them could change under metamorphosis. Some metamorphosis scenarios involve cases where the app ID remains unchanged, and updates are pushed to users who have already installed the app. The behaviour, branding and functionality changes can come as a total surprise to the user. Also, at Google Play Store, potential new users will not see any transformations an app has undergone and are compelled to \emph{``download an app as you see it"}, even if we remember \emph{``what it used to be"}. We next discuss and present further results to emphasise why app metamorphosis is an important phenomenon that requires attention in app markets.


\subsection{App counterfeits and plagiarism}
In two instances out of 88 potential re-birth instances that were flagged by our methodology, we observed that the original app was discontinued and re-born with a different app ID. Further analysis established that the original apps did not discontinue, instead they went through some major changes. 

For example, Pi Music Player was renamed to Pi Music Player- MP3 Player with drastic changes in the app icon and other metadata as portrayed in Fig.~\ref{fig:sec_risks}(a). Noticing this change, a counterfeit app emerged in the name of Music Player with the same app icon that used to be with the authentic app and reached nearly 100k+ downloads. Similarly, a counterfeit app called Truck Simulator appeared when the original app Truck Simulator USA was changed to Truck Simulator USA Evolution, as shown in Fig.~\ref{fig:sec_risks}(b). This counterfeit app, however, is now removed from Google Play Store but managed to reach $\sim200k$ downloads. 

In one instance of app re-branding, when Google Duo changed to Google Meet, another app developer picked up an almost similar app icon to the discontinued Google Duo app, aiming to trick users into downloading the app based on icon familiarity. This is shown in Fig.~\ref{fig:sec_risks}(c). Another instance, Fig.~\ref{fig:sec_risks}(d), we observed is a developer named Baby Penguin maintaining an app visually similar to the popular imo-International Calls \& Chat app, potentially victimising more than 10 million users, until it was discontinued/banned from Play Store. 



It is important to note that we were not actively looking for counterfeits as they are outside our main research focus. Rather, we found them to be a byproduct of our metamorphosis analysis. Nonetheless, results like these show the security and privacy risks even to the most tech-savvy users in identifying the authenticity of an app among the top apps in Google Play Store. 
The examples we observed did not contain any malware indications when evaluating their installation files with VirusTotal service, rather observed as an effort to tap on to existing popular apps in top 10,000 category.
While previous work has developed methods for app counterfeit detection when both the target and the counterfeit apps are in the market and highlighted the associated associated risks, such as counterfeit apps being channels for malware distribution~\cite{karunanayake2020multi}, our metamorphosis analysis provides insights into how other scenarios lead to counterfeits. 

We encourage app market operators to deploy a similar pipeline as we proposed and actively investigate new additions to the app market. As emphasised in Appendix~B, search efficiency improvement of our method further enables it to be deployed in the real-world scale where an app is introduced every 1.13 minutes to the PlayStore~\cite{42google}. We suggest the app markets to observe the resultant matching app(s) if existing and bin the results to appropriate metamorphosis categories we identified. Ideally, the operators can perform this for multiple past-indexed years to obtain an even finer-granular picture. The market operators have a further advantage to identify if a matching app is counterfeit or not as they have the relevant history for a particular change (e.g. the developer changed the name, the developer re-branded the app, etc.) and further investigate any other potential red flags. Though it is out of our scope, malware detection techniques (e.g. VirusTotal) and static or dynamic code analysis techniques are resource intensive and could not be done for every new app. Instead, any red flags from our method are ideal to be selected for such additional checks. We believe the metamorphosis concept we propose simplifies finer-granular investigation steps for market operators. On the contrary, the researchers and end-users hardly have information on the history of an app or an app-developer unless it is covered by media such as with music.ly to tiktok example.

\subsection{Permission changes}

\begin{figure}[ht]
    \centering
    \includegraphics[width=0.49\textwidth]{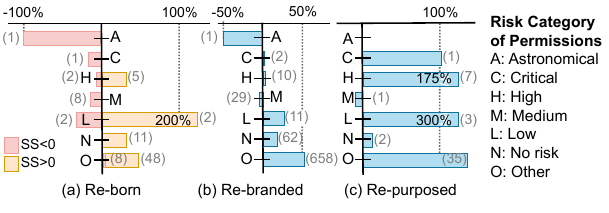}
    \caption{Percentage change of permissions according to the risk category when the apps underwent three metamorphosis categories. (x) indicates the number of permissions that changed. }
    \vspace{-0.2cm}
    \label{fig:permission_change}
\end{figure}

We evaluated the changes in the permissions of the manually verified apps belonging to re-birth (74), re-branding (215) and re-purposing (8) metamorphosis categories and their percentage changes are shown in Fig.~\ref{fig:permission_change}. 
We categorised each permission based on the risk associated with it as previously suggested by~\cite{permission_risks}. 

When an app undergoes a re-birth, it is essentially a new app; therefore, a change of permissions is expected. Among the 74 apps, we observe that astronomical and critical permission requests have been reduced. Re-branded apps and re-purposed apps, on the other hand, are essentially the same apps if the user installed them before the transition period as there is no change in app ID and the app will be updated automatically (based on the user settings) and new permission requests will be prompted to the users. 

With re-branding, we do not observe significant permissions growth for medium and above risk categories. Comparatively, the re-purposed apps have a significant growth of permission requests for high and critical. It is questionable whether existing users are transparently informed about when apps undergoing major changes that re-purpose their core functionality, as users tend to routinely allow permission prompts without much scrutiny. Therefore, we reiterate the importance of users remaining vigilant about routine permission requests from apps and carefully evaluating their necessity.  

From a privacy standpoint, majority of end-users are expressing concerns about data collection and retention practices, but there is an ambiguity surrounding how sensitive information is handled during app metamorphosis. While Google Play places a strong emphasis on developers disclosing their practices, specific instances of app metamorphosis, such as when an app changes its developer (app-transferred), are not explicitly addressed on the Google Play Store. Consequently, most app users are unlikely to notice such changes and remain unaware of how their information is being passed along. Therefore, it is crucial to carefully observe privacy change notices when they appear while using an app.

Similarly, app market operators should also implement mechanisms to detect changes in app permissions that may occur during legitimate app metamorphosis. When significant changes are identified, such as changes in authorship or a large number of new permissions, users should be notified promptly. These notifications should clearly inform users that the app has undergone substantial modifications and that caution is advised before continuing to use it. 
\vspace{-0.3cm}

\subsection{Breach of user trust}

Developers are encouraged to provide routine updates to the apps and as a result, they have the freedom to change app matadata to reflect such changes. In fact, there are apps that can benefit from changing the theme according to the season, such as snowy themes during the winter tri-month (e.g., shopping apps and games providing holiday-themed in-app purchases). However, due to the freedom of metadata change, we observed that the genre category (ref. Sec.~VII.E), content rating category (ref. Sec.~VII.F) and even the revenue model (ref. Sec.~VII.G) have been changed while the app was not undergoing major reformations in functionality or appearance. We have identified several factors that could have influenced these changes, yet this creates a breach in user trust. A game could change the content rating from PG (parental guidance) to M (mature), however, a child who downloaded the game previously could be exposed to sensitive content not intended for them, and often the developers could collect blind consenting in the form of “We are updating the app” report that users rarely read. Similarly, a game that changed from “puzzle” into “simulation” where the original users expected it to be “puzzle” is again a breach of their trust and a waste of their time. An app becoming free after being a paid app should consider having refund policies for their existing users, yet most of such apps being one-time payments at the beginning makes it nearly impossible to refund. In such a context, it is important that app market operators be more vigilant about these metamorphosis categories and promptly note their existing user base about these particular changes, in addition to what the developers note.
\vspace{0.4cm}

%% file: sections/8_Conclusion.tex
\section{Concluding Remarks}
\label{Sec:Conclusion}
Using our proposed multi-modal app search methodology, we investigated the phenomenon of \emph{``app metamorphosis''} occurring on the Google Play Store between two snapshots taken five years apart. To the best of our knowledge, this is the first study examining this phenomenon. By defining a success score (SS) for each app, we quantitatively characterised the metamorphosis categories. Our observations revealed that although these apps do not follow the traditional app life cycle of incremental updates, the majority of apps in these categories are more successful compared to an average app within the top 10,000 category in 2018.


We found interesting forms of metamorphosis such as re-birthing (average ${SS}\approx-15\%$), re-branding (average ${SS}\approx21\%$),  re-purposing (average ${SS}\approx14\%$), for which we provided details on identification and analysis. In addition to these we also found other forms of metamorphosis such as genre changes, content-rating changes, revenue model changes, developer transfers, targeting multiple demographics, and coming in multiple versions. We observed that re-branding is more successful than the other categories.

%


While being restricted to only two snapshots is a limitation, our focus was on identifying significant cases of 'metamorphosis' over a period of five years. Introducing an intermediate snapshot could shift the focus toward smaller, incremental changes, which deviate from our primary goal. For example, re-branding is often a gradual process involving changes such as app icons, descriptions, or minor UI adjustments. An intermediate snapshot might introduce ambiguity between seasonal re-branding, incremental re-branding, and full re-branding, complicating detection. Similarly, for re-birth scenarios, where a re-born app co-exists with its old version temporarily, an intermediate snapshot could blur the classification, making it unclear whether to treat this as a re-birth or a different version entirely. However, we were still able to conclusively validate the majority of our results. For example, out of all potential rebirth and re-branding results, we manually verified approximately 85\% and 78\%, respectively, to be conclusive and strong examples. 

Additionally, we presented insights into privacy and security risks arising from metamorphosis, such as app counterfeiting, requests for additional permissions, and changes in app behaviour. Our results show that app ecosystems need to pay close attention to the metamorphosis phenomenon to maintain the integrity of their stores. Based on our proposed methodology, market operators can analyse app metadata changes in a similar way how app users perceive information prior to installation, and based on observable metamorphosis categories (if any), they can further analyse latest versions against possible past versions and flag any mismatches. Market operators have much further capacity to view the history of app developers, perform static and dynamic code analysis to further evaluate such flagged instances. The flags created by our method will aid in narrowing down this extensive analysis process to the metamorphosis apps that could be potentially risky and operators will have the capacity to take them down even before user complaints.

%% file: sections/9_Appendix.tex



\section{Appendix}
\label{sec: appendix}

\begin{table*}[b]
  \scriptsize
  \caption{\centering Harmonic mean on the validation set for different embedding combinations and voting thresholds}
  \label{tab:ablations_modality_sele}
  \centering
  \begin{tabular}{lrrrrrrrrrrrr}
    \cline{2-13}
    & \multicolumn{2}{c}{$\alpha$ = 1} & \multicolumn{2}{c}{$\alpha$ = 2} & \multicolumn{2}{c}{\textbf{$\alpha$ = 3}} & \multicolumn{2}{c}{$\alpha$ = 4} & \multicolumn{4}{c}{Harmonic mean} \\
    \cline{2-13}
    &      &  No     &     & No      &      &  No     &      & No      &       &&    $\alpha$   &       \\
    \cline{10-13}
    & Match & Match &  Match & Match & Match & Match &  Match & Match &    1   &   2    &3&    4   \\
    \hline
    $MPNet_{desc} (M)$  & 93.0&0.0& -& -& -& -& -&	-& 0.0& -& -& -	\\
    $TFIDF_{name} (T)$& 89.6&0.0& -& -& -& -& -&	-& 0.0& -& -& -	\\
    $ViT_{icon}$ & 75.6&0.0& -&	-& -&	-&	-&	-& 0.0& -& -& -\\
    $VGG19^c_{icon}$ & 70.4&0.0& -&	-& -&	-&	-&	-&	0.0& -&	-&	-\\
    $StyTr^{2^c}_{icon}$ & 58.4& 0.0&	-&	-& -&	-&	-&	-&	0.0& -&	-&	-\\
    $VGG19^s_{icon}$ & 74.4&0.0& -&	-& -&	-&	-&	-&	0.0& -&	-&	-\\
    $StyTr^{2^s}_{icon}$ & 62.4& 0.0&	-&	-& -&	-&	-&	-&	0.0& -&	-&	-\\
    \hline
    $M + T$ & 88.8&0.0&	83.4&	56.8&	73.6&	92.6 &	-&	-&	0.0 & 67.6 & 82.0 & -\\
    $M + ViT_{icon}$ & 91.8&0.0&	70.6&	84.0&	63.0&	95.6&	-&	-&	0.0& 76.7&75.9&		-\\
    $M + VGG19^c_{icon}$ & 87.6&0.0&	68.0&	92.6&	60.8&	96.2&	-&	-& 0.0& 78.4&74.5&		-\\
    $M + StyTr^{2^c}_{icon}$ & 89.2 & 0.0 & 56.2 & 96.0 & 50.2 & 98.2 &	-&	-&	0.0& 70.9&66.4& -\\
    \hline
    $M + ViT_{icon} + T$ & 90.2& 0.0 &	88.6 & 45.4 &	84.8 & 89.0 & 56.0 & 97.6 &	0.0 & 60.0 & 86.8 & 71.2\\
    $M + VGG19^c_{icon} + T$ & 89.4 & 0.0 & 88.6 &	43.8 & 84.2 & 89.0 & 52.8 & 98.0 & 0.0 & 58.6 & 86.5 & 68.6\\
    $M + StyTr^{2^c}_{icon} + T$ & 88.4 & 0.0 & 86.8 & 45.6 & 81.8 & 90.8 & 43.0 & 98.8 & 0.0 & 59.8 & 86.1 & 59.9\\
    \hline
    $M + T + ViT_{icon} +VGG19^s_{icon}$ & 88.8 & 0.0 & 88.4 &	32.2 & 86.2 & 85.6 & 70.2 & 95.6 & 0.0 & 47.2 & 85.9 & 81.0 \\
    $M + T + ViT_{icon} +StyTr^{2^s}_{icon}$ & 90.0 & 0.0 & 89.4 & 35.4 & 86.2 & 86.0 & 65.8 & 96.4 & 0.0 & 50.7 &	86.1 & 78.2 \\
    $M + T +VGG19^c_{icon} +VGG19^s_{icon}$ & 86.2 & 0.0 & 85.8 & 28.8 & 83.6 & 86.6 & 64.4 & 96.2 & 0.0 & 43.1 & 85.1 & 77.2 \\
    $M + T +VGG19^c_{icon} +StyTr^{2^s}_{icon}$ & 89.2 & 0.0 & 88.8 & 37.6 & 85.4 & 88.2 & 62.6 & 97.2 & 0.0 & 52.8 & 86.8 & 76.2 \\
    $\bf M + T +StyTr^{2^c}_{icon} +VGG19^s_{icon}$ & 88.8 & 0.0 & 88.4 & 36.8 & 86.0 & 88.2 & 63.0 & 97.0 & 0.0 & 52.0 & \textbf{87.1} & 76.4\\
    $M + T +StyTr^{2^c}_{icon} +StyTr^{2^s}_{icon}$ & 89.2 & 0.0 & 88.8 & 37.6 & 85.4 & 88.2 & 62.6 & 97.2 & 0.0 & 37.1 & 84.5 & 69.7 \\
    \hline
  \end{tabular}
\end{table*}

We discuss the ethical considerations of our work in Appendix~\ref{sec:ethics}, and in Appendix~\ref{sec:search_eff}, we provide the comparison of search efficiency between our model and Karunanayake et.al~\cite{karunanayake2020multi}. Finally, in Appendix~\ref{sec:val_app_dist} and Appendix~\ref{sec:repurAS} we describe app distribution of the validation set and hyper-parameter selection for re-purposed apps.


\subsection{Ethics}
\label{sec:ethics}

During the crawling stages, we only crawled publicly available app metadata hosted in Google Play Store, and to avoid any disturbances to its operation, we visited Play Store pages at a very slow rate. Beyond that, we did not process any data related to app users, such as app reviews.

\subsection{Search efficiency}
\label{sec:search_eff}

\begin{figure}
    \centering
    \includegraphics[width=0.49\textwidth]{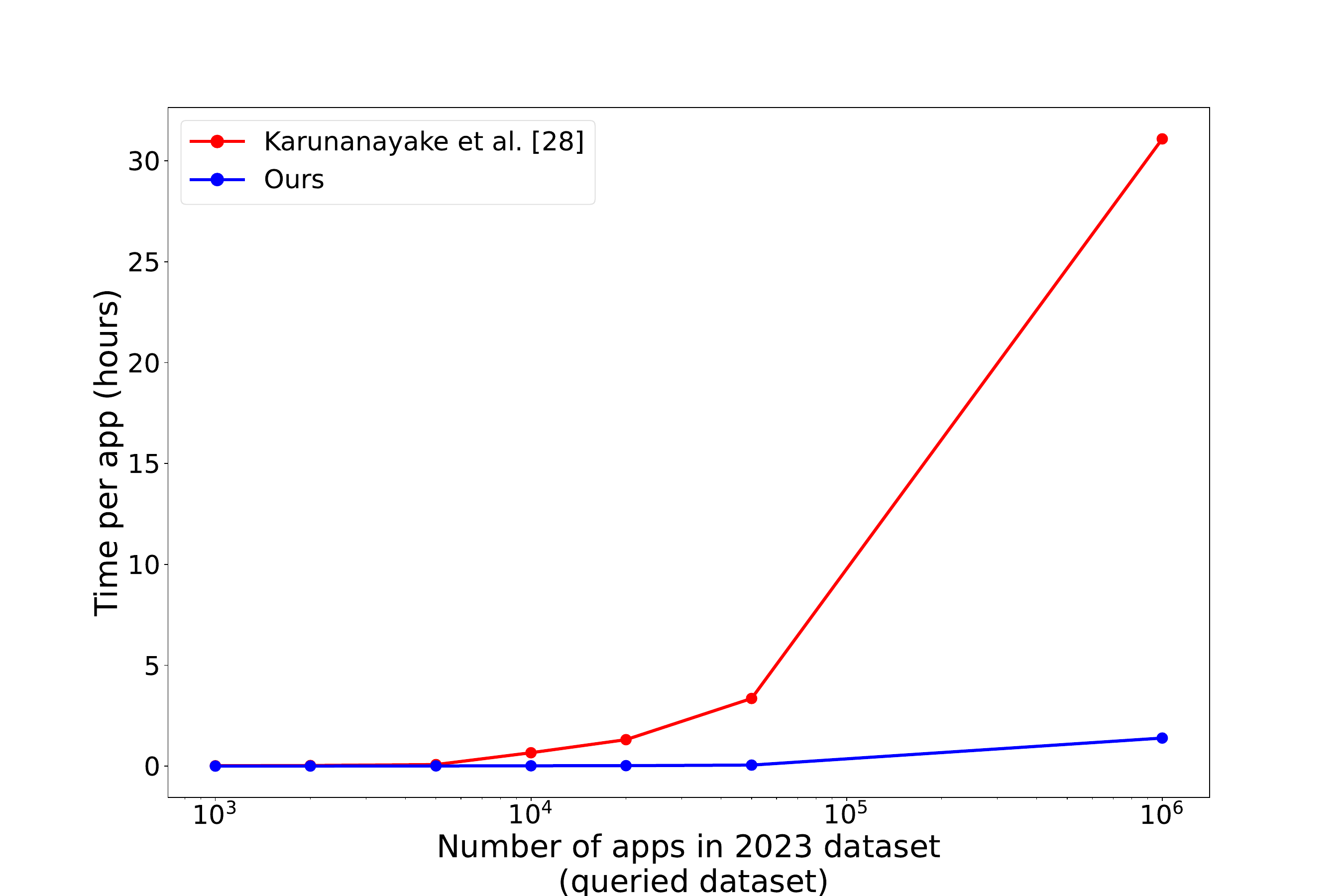}
    \caption{Comparison of Time Efficiency: Proposed Method vs. Karunanayake et al.~\cite{karunanayake2020multi}}
    \label{fig:efff_search}
\vspace{-0.3cm}
\end{figure}

Here, we show the comparison of the efficiency of our proposed method with the method introduced by Karunanayake et al.~\cite{karunanayake2020multi}. Both methods rely on generating embeddings, but their subsequent processes differ significantly. The method~\cite{karunanayake2020multi} requires an extensive grid search to identify optimal parameters, where each parameter ranges from 1 to 9. Using our validation set (Sec.~III), this hyper-parameter tuning took approximately 17.25 hours (62,097.613 seconds). However, this step is performed only once. After hyper-parameter tuning,~\cite{karunanayake2020multi} performs a similarity search using optimised parameters. In contrast, our method generates FAISS indices and uses those indices to find the most similar counterpart. We tested this on the test set (Sect.~III) by progressively increasing the size of the queried dataset. Fig.~\ref{fig:efff_search} shows the time taken to find a counterpart for a single app across increasing dataset sizes. The time taken by~\cite{karunanayake2020multi} increases exponentially as the queried app space grows. In contrast, our method scales more efficiently, demonstrating near-linear growth. This makes our approach significantly more practical for larger datasets, where~\cite{karunanayake2020multi} becomes infeasible due to the exponential increase in processing time.

\begin{figure}
    \centering
    \includegraphics[width=0.45\textwidth]{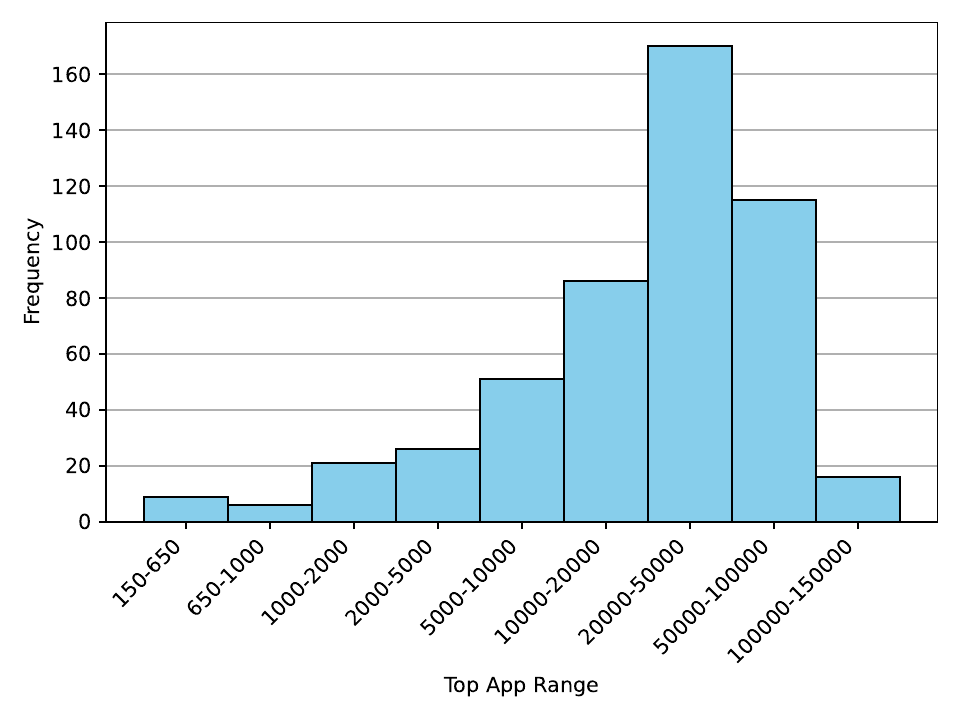}
    \caption{Distribution of the validation set across dataset}
    \label{fig:vld_set_D}
\vspace{-0.3cm}
\end{figure}

\begin{figure}
    \centering
    \includegraphics[width=0.48\textwidth]{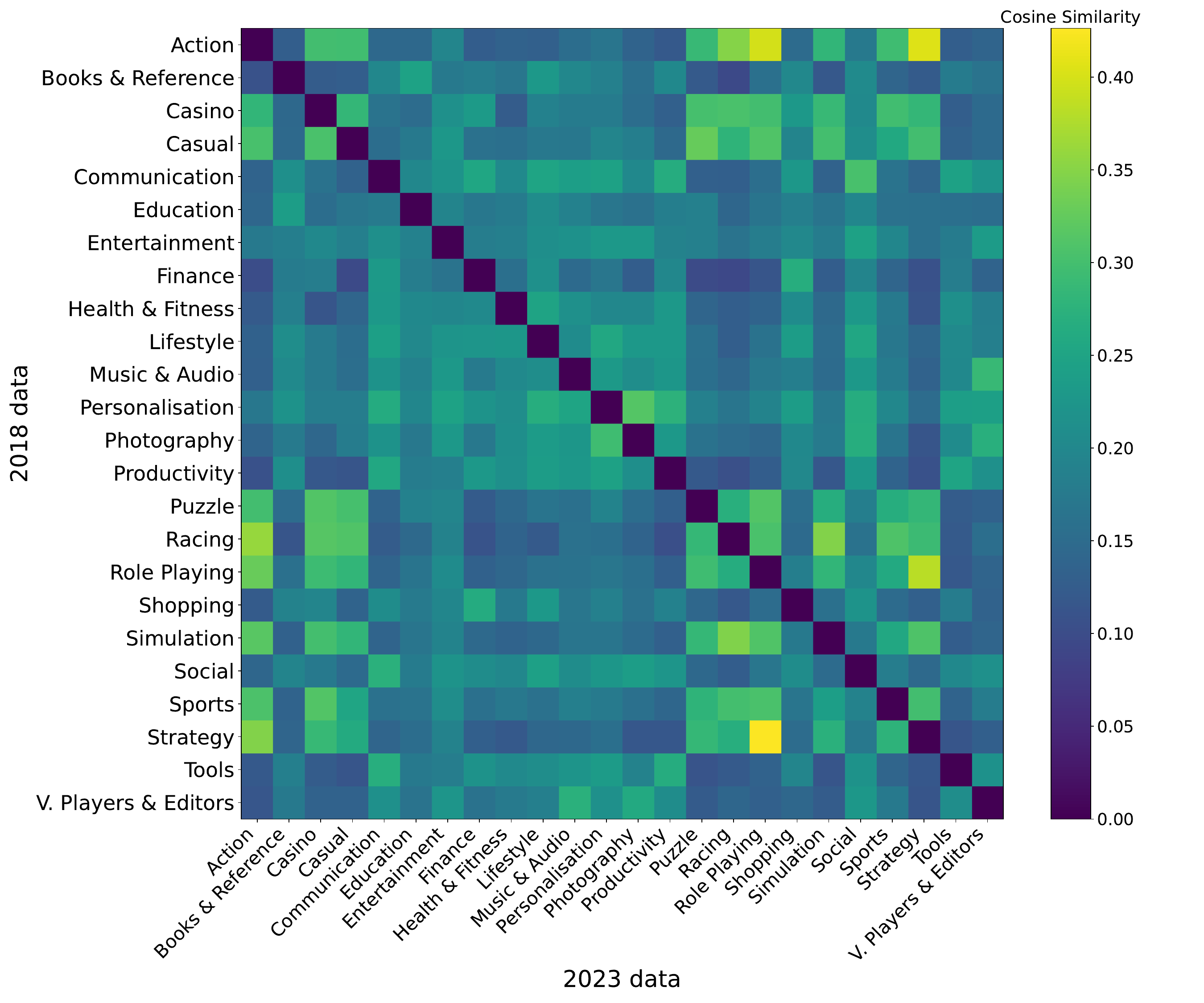}
    \caption{Average cosine similarity between app genres from two datasets 2018 and 2023.}
    \label{fig:repur_heat}
\vspace{-0.3cm}
\end{figure}

\begin{figure}
    \centering   
    \includegraphics[width=0.45\textwidth]{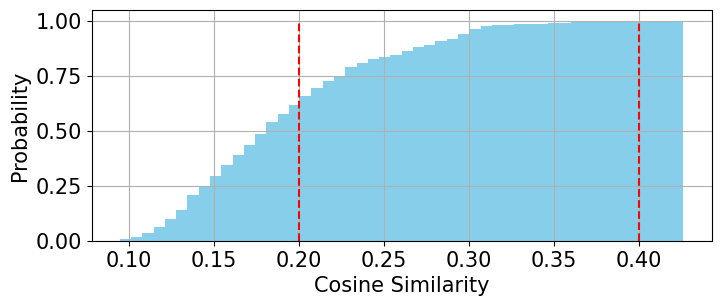}
    \caption{CDF of average cosine similarity between app genres from two datasets 2018 and 2023. }
    \label{fig:repur_cdf}
\end{figure}

\subsection{Validation app distribution}
\label{sec:val_app_dist}
As shown in the Fig.~\ref{fig:vld_set_D}, app selection in the validation set ranges from top-150 to top-150k allowing to select a generic value for the hyper-parameter $\alpha$.

\subsection{Modality selection}
\label{sec:modality_sel}
To evaluate the effectiveness of each modality, we conducted experiments on the validation set described in Sec.~III. These experiments involved varying the number of modalities used for similarity matching and calculating the harmonic mean of accuracy for both matches and non-matches across four different occurrence count thresholds ($\alpha$ values). Starting with a single modality, we incrementally increased the number of modalities up to five. The findings are depicted in the Tab.~\ref{tab:ablations_modality_sele}.  The results demonstrate that the selected combination of four modalities: MPNet app description  embeddings ($MPNET_{desc}$ or $M$), TF-IDF app name and developer name embeddings ($TFIDF_{name}$ or $T$), VGG19 style embedding ($VGG19^s_{icon}$) and $StyTr^2$ content embeddings ($StyTr^{2^c}_{icon}$) for app icon achieves the highest performance. This analysis supports our claim that the chosen combination is the most effective for identifying matches and no-matches.

\subsection{Ablation study - re-purposed apps}
\label{sec:repurAS}



Having core functionality changes in re-purposed apps, we highlighted that the app description embeddings should deviate significantly from one another. To validate the choice of the $0.2$–$0.4$ similarity threshold, we conducted an experiment comparing the cosine similarity of app descriptions across genres in the 2018 and 2023 datasets with the intuition that dis-similar genres are likely to contain dis-similar descriptions explaining the functionality. In this experiment, for each genre, 100 apps were sampled based on availability, and their average cosine similarity of app description embeddings were computed and portrayed in Fig.~\ref{fig:repur_heat} and Fig.~\ref{fig:repur_cdf}. The average inter-genre cosine similarity is $0.193$ and the maximum cosine similarity is $0.43$. Furthermore, we observed apps with descriptions in non-English languages producing similarities around $0.15$, contributing to noise. To mitigate this and focus on apps that demonstrate significant but interpretable functional shifts, we selected a threshold of $0.2$–$0.4$. This range effectively excludes noisy data.


%% file: Main_TMC.bbl
\begin{thebibliography}{10}
\bibitem{ashgoogle}
{Ash Turner}, ``The rise of android: Why is android successful?.''
\end{thebibliography}
